\documentclass{article}

\usepackage{PRIMEarxiv}
\usepackage[utf8]{inputenc} 
\usepackage[T1]{fontenc}    
\usepackage{hyperref}       
\usepackage{url}            
\usepackage{booktabs}       
\usepackage{amsfonts}       
\usepackage{nicefrac}       
\usepackage{microtype}      
\usepackage{lipsum}
\usepackage{fancyhdr}       
\usepackage{graphicx}       
\usepackage{amsmath}
\usepackage{bm}
\usepackage[utf8]{inputenc}
\usepackage{textcomp}
\usepackage{subcaption}
\usepackage[version=4]{mhchem}
\usepackage{siunitx}
\usepackage{tabularray}
\usepackage{longtable,tabularx}
\usepackage{rotating} 
\usepackage{xcolor}
\usepackage{verbatim}
 
\usepackage{amssymb}
\usepackage{subcaption}
\usepackage{caption}
\usepackage{float}
\usepackage{booktabs}
\usepackage{listings}
\usepackage[ruled]{algorithm2e}

\graphicspath{{media/}}     

\title{UQ of 2D Slab Burner DNS: Surrogates, Uncertainty Propagation, and Parameter Calibration 
}

\author{
  Georgios Georgalis, Alejandro Becerra \\
  Tufts Institute for AI and Department of Mathematics\\
  Tufts University \\
  Medford, MA 02155, USA\\
  \texttt{\{georgios.georgalis, alejandro.becerra\}@tufts.edu} \\
   \And
  Kenneth Budzinski, Matthew McGurn, Danial Faghihi, Paul E. DesJardin \\
  Department of Mechanical and Aerospace Engineering \\
  University at Buffalo, The State University of New York, \\
  Buffalo, NY 14260, USA\\
  \texttt{\{klbudzin, -, danialfa, ped3\}@buffalo.edu} \\
   \And
   Abani Patra \\
   Tufts Institute for AI and Department of Mathematics \\
   Tufts University \\
   Medford, MA 02155, USA \\
   \texttt{abani.patra@tufts.edu} \\
}

\begin{document}

\maketitle
\thispagestyle{fancy}

\begin{abstract}
The goal of this paper is to demonstrate and address challenges related to all aspects of performing a complete uncertainty quantification analysis of a complicated physics-based simulation like a 2D slab burner direct numerical simulation (DNS). The UQ framework includes the development of data-driven surrogate models, propagation of parametric uncertainties to the fuel regression rate--the primary quantity of interest--and Bayesian calibration of the latent heat of sublimation and a chemical reaction temperature exponent using experimental data. Two surrogate models, a Gaussian Process (GP) and a Hierarchical Multiscale Surrogate (HMS) were constructed using an ensemble of 64 simulations generated via Latin Hypercube sampling. HMS is superior for prediction demonstrated by cross-validation and able to achieve an error $<15 \%$ when predicting multiscale boundary quantities just from a few far field inputs. Subsequent Bayesian calibration of chemical kinetics and fuel response parameters against experimental observations showed that the default values used in the DNS should be higher to better match measurements. This study highlights the importance of surrogate model selection and parameter calibration in quantifying uncertainty in predictions of fuel regression rates in complex combustion systems.
\end{abstract}

\section{Introduction}
Hybrid rocket systems, characterized by a fuel and oxidizer in two different states, are considered better alternatives compared to other propulsion systems due to their ability to operate with a highly dense fuel source similar to a bi-propellant solid motor but with the operational advantages of a liquid motor \cite{MAZZETTI2016286, Karabeyoglu98, OKNINSKI2021260}. Many experimental studies have been conducted that test slab burner hybrid rocket setups with high alkane fuels such as paraffin because these fuels achieve higher regression rates compared to other options (e.g., see \cite{CARRICKFUELS, Georgalis_2023, SURINA2022160, Kobaldrheological, GLASER2023186, Karabeyoglu_Exp}). The higher regression rates result from the combustion phenomena of such fuels: the formation of a liquid layer on the surface of the solid, which, together with the formation of instabilities in the fuel-oxidizer interface, leads to the entrainment of combustible liquid droplets into the main flow \cite{KARABEYOGLU1}. Despite the observation of these concepts experimentally, creating accurate computational models for hybrid rockets is very challenging due to the complexity of the physics involved: flow transport, atomization, turbulent combustion, and complex chemical kinetics need to be solved together. 
Accurate prediction of reacting flows, in general, is a challenge in predictive modeling because of the multiscale nature of combustion, as the underlying coupled physical phenomena occur under a wide range of time and length scales. Namely, elementary chemical reaction processes happen in short time scales (e.g., chain reactions that occur on the order of $10^{-10}s$), whereas the scales of the flow transport are much larger on the order of $10^{-4}$ to $10^{-2}s$ ~\cite{Echekki_2009}. Similarly, in terms of length dimension scales, the flame structures under radiation and soot effects are smaller than physical scales of flow transport in a combustion chamber ~\cite{PETERS20091,MSmodelingcomp}. Apart from combustion being multiscale, there exists varying degrees of coupling between the physical processes involved. For example, chemical reactions and molecular diffusion are tightly coupled at fine scales \cite{POPE20131} and need to be modeled e.g., via reduced-order flamelet manifolds (see \cite{KennyScitech, Amolscitech, MUELLER2020287, PERRY2022112286}) or stochastically with probability distributions \cite{HAWORTH2010168,PEI20152006} in turbulent combustion simulations since grids chosen to capture flows are not sufficiently fine. 

These challenges of the coupled multiscale phenomena have led to the development of various computational approaches, such as Direct Numerical Simulation (DNS). In DNS, the Navier-Stokes equations are coupled with the thermochemical species equations and solved numerically on a grid that sufficiently captures the underlying scales \cite{DESJARDIN1996343}. 
An issue that arises when using DNS solvers is that the computational cost is often prohibitive for any detailed data-driven analysis that requires many simulation runs like uncertainty quantification (UQ). DNS simulations of combustion are often not standalone components, but parts of a larger framework that may include experiments and/or other models that either provide inputs to the DNS or use the DNS outputs. In this context, a careful UQ analysis on the DNS is required to either validate the DNS solvers themselves for decision-making depending on the application or to be able to propagate uncertainty through the DNS to other models. Any standard forward UQ or calibration analysis requires large ensembles of model evaluations to adequately approximate the true posterior distribution of the output quantities of interest (QoIs) in forward UQ or the posterior of the calibration parameters \cite{SHIRIAN2023106046, JANSSEN2013123}, which is infeasible given the DNS computational cost. Therefore, it is common practice and a requirement to create low-cost and high-accuracy surrogate models (i.e., emulators, approximate models) \cite{grammacytext,KennedyOhagan, sackswelch} for uncertainty quantification and calibration of these expensive solvers. 

There are many methods for developing surrogate models. Polynomial chaos expansions (PCE) approximate the input to QoI function $\mathcal{M}:\boldsymbol{x}\in \mathcal{D}_{\boldsymbol{X}}\subset \mathbb{R}^{M}\mapsto \boldsymbol{y}=\mathcal{M}(\boldsymbol{x})$ with polynomial approximations \cite{PCElit}. Canonical low-rank tensor approximations (LRA) approximate the input to QoI function with products of univariate functions that can also be orthogonal polynomials similar to PCE \cite{LRAlit}. Feedforward neural networks map inputs to QoIs via learning the weights and biases between layers of neurons \cite{CHAN2018493, singh2024framework}. Kriging or Gaussian Processes (GPs) assume the likelihood function of the QoI to be a multivariate normal distribution modeled as the addition of a mean regression term and a Gaussian stationary random process \cite{KrigingGP}. Combinations of these methods are also options for surrogate models, e.g.,  Deep GPs \cite{RADAIDEH2020106731}, where the data between layers of a neural network are modeled as a multivariate GP. Traditional surrogate methods such as the ones we outline above often perform poorly when used in multiscale problems, because these problems include characteristics that are computationally challenging for these basic methods. Firstly, multiscale problems have a very large input space and that requires specific techniques to address the curse of dimensionality before proceeding with surrogate models \cite{ 10.1007/11494669_93,hou2022dimensionality}. Secondly, methods such as GPs are not scalable because they include covariance matrix inversions, which are computationally expensive. Thirdly, the assumptions of smoothness and stationarity imposed by Gaussian assumptions or polynomials does not necessarily hold for multiscale problems. Lastly, the ability of simpler methods to capture non-linear trends from high-dimensional data is limited \cite{10.5555/927743}. 

The main goal of this paper is to systematically address challenges related to all aspects of performing a complete UQ analysis of a complicated physics-based simulation like a 2D slab burner DNS. We provide insights related to the development of an acceptable surrogate model, the propagation of uncertainty of the DNS inputs to the QoI, and the calibration of DNS parameters based on observations of experimental data. The DNS used in this work is ABLATE (Ablative Boundary Layers At The Exascale), an in-house DNS solver developed for simulating a multiphase 2D slab burner setup to better understand combustion in hybrid rocket motors, which includes ablation, radiation, and soot modeling\cite{BUDZINSKI2020248,DESJARDIN1996343,KennyScitech}. ABLATE uses the PETSc libraries from Argonne National Laboratory \cite{petsc-user-ref} for data management, and parallel large scale computation. For the work described in this paper, ABLATE simulated the 2D reacting flow between pure oxygen $O_2$ (the oxidizer) and Methyl methacrylate MMA (the fuel) with chemistry modeled by a reduced mechanism. To accurately capture the multiscale behavior and overcome the computationally prohibitive UQ analysis of the DNS simulations, we build both GPs and a novel computationally efficient hierarchical multiscale surrogate (HMS) \cite{Shekhar2020b}. HMS is based on a forward-backward greedy approach to construct a set of data driven basis functions with a multiscale structure. Forward-backward here means the iterative selection of basis functions at different scales. This approach generates  hierarchical basis functions belonging to Reproducing Kernel Hilbert Spaces (RKHS), and the hierarchical models are able to accurately represent irregularly structured data at different resolutions (scales), rapidly minimizing the error in fitting by using kernels at successively finer scales. We conduct a forward UQ analysis using the developed surrogate and uncertainty in a subset of inputs identified using sensitivity analysis and propagating the uncertainty of the DNS inputs to the slab burner output QoI. The inputs to the DNS are the velocity factor for the far-field inlet oxidizer flux assuming a parabolic profile ($V_{in}$), the thermochemical properties of the chemistry mechanism ($\theta_{c}$: activation energies $E_*$, temperature exponents $b_*$, and pre-exponential factors $A_*$), the fixed geometry of the slab burner, and the fuel parameters ($\theta_{f}$). Combustion DNS solvers produce a variety of outputs that may be of interest depending on the goals. Here, we are most interested in QoIs along the boundary of the slab, namely the regression rate ($\dot{r}$). The regression rate has a direct impact on the geometrical design of the rocket motor and its performance (\cite{Zilliac_Karabeyoglu_AIAA_RR}, \cite{Zilliac-uq}). Lastly, we leverage the constructed surrogate model and experimental regression rate measurements to calibrate the fuel ($\theta_f$) and chemistry ($\theta_c$) parameters of the DNS. In particular, we use the Bayesian method to infer the probability distribution of the latent heat of sublimation ($l_v$) and activation energy for a particular important reaction ($E_1$).

The paper is organized as follows. Section 1 acts as the introduction and provides motivation for uncertainty quantification of hybrid rocket DNS solvers. Section 2 presents ABLATE, the DNS solver for a 2D slab burner with descriptions of its inputs and outputs. Section 3 introduces the hierarchical multiscale surrogate (HMS) and benchmarks, as well as the forward UQ and calibration methods. Section 4 includes results and discussions for the development of the surrogate models, forward UQ, and calibration. Section 5 is the concluding section with directions for future work.

\section{ABLATE: Coupled Flow/Combustion DNS}
In this section, we describe our coupled flow/combustion solver ABLATE (\href{https://ablate.dev/}{https://ablate.dev/}), its inputs and outputs, and how we used it to generate the dataset to train the surrogates in Section 3. The solver simulates a 2D slab burner setup for the combustion of MMA as the fuel and pure oxygen $O_2$ as the oxidizer. 

\begin{figure}
\centering
\includegraphics[width=.7\textwidth]{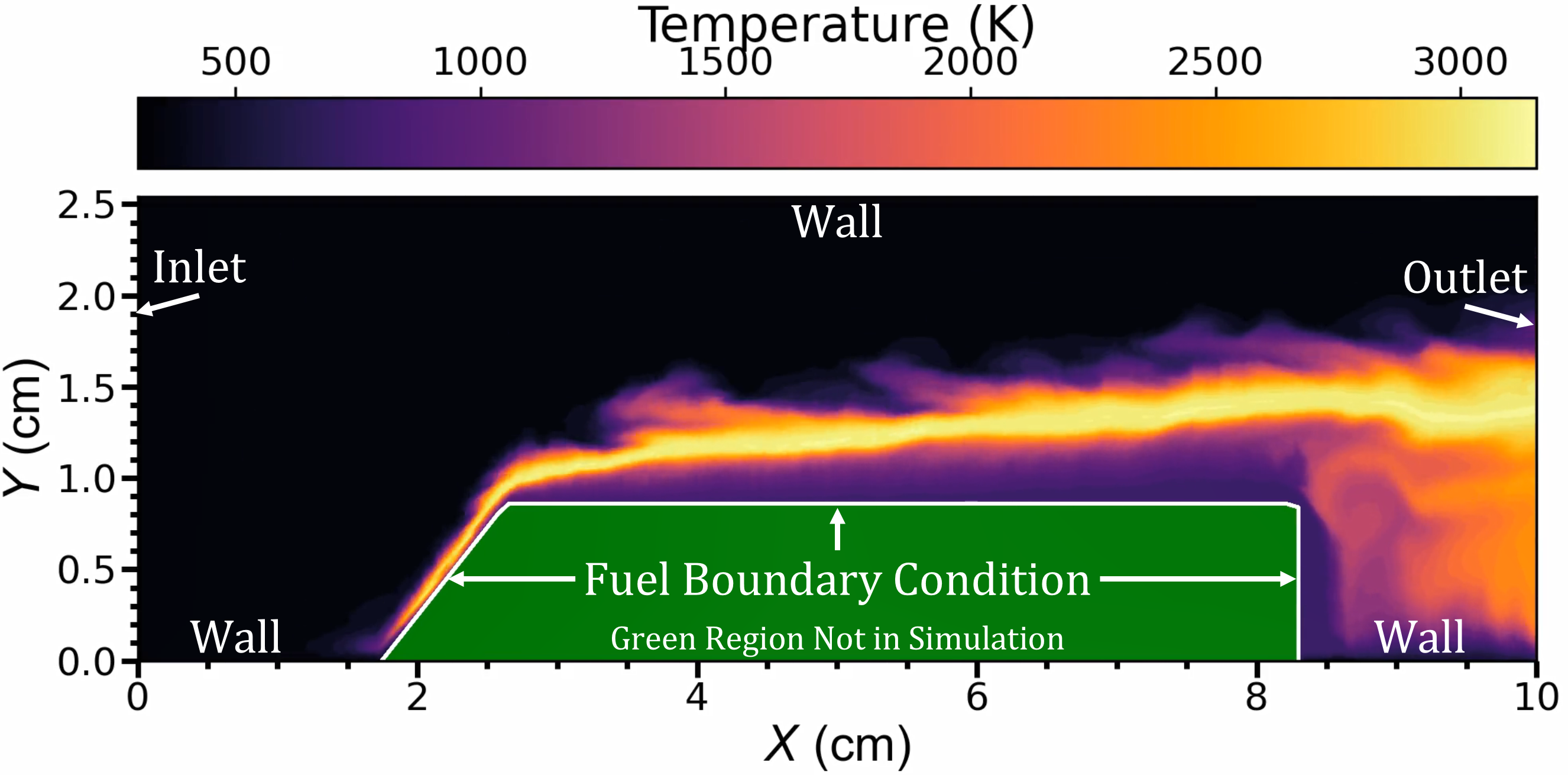}
\caption{A sample temperature contour of the computational domain showing the specified inlet, outlet, isothermal wall, and fuel boundaries. The QoI is the time averaged regression rate at each point along the fuel boundary.}
\label{fig:SlabLabeled}
\end{figure}

Figure ~\ref{fig:SlabLabeled} shows the two-dimensional slab burner domain. An inlet of pure oxygen flows from the left side over the fuel slab where it reacts with the fuel vapors and flows through the outlet of the slab burner on the right side. There are four different boundary conditions used in the simulation depending if the boundary is a (1) wall, (2) inlet, (3) outlet, or (4) the fuel.
The gas phase system for the slab burner environment is described by the reacting compressible Navier-Stokes equations coupled with radiation heat transfer:
\noindent
\begin{align}
    \label{eqn:NS}
&\frac{\partial \rho }{\partial t} + \vec{\nabla}\cdot ( \rho \vec{u}) = 0\nonumber\\
&\frac{\partial \rho \vec{u}}{\partial t} + \vec{\nabla} \cdot (\rho \vec{u} \vec{u}) = - \vec{\nabla}P + \vec{\nabla}\cdot\bm{\tau}^T  \nonumber\\
&\frac{\partial (\rho Y_k )}{\partial t} + \vec{\nabla} \cdot (\rho \vec{u} Y_k ) = \vec{\nabla}\cdot\left( \rho D\vec{\nabla} Y_k\right)+ \dot{m}'''_k   \\
&\frac{\partial (\rho E )}{\partial t} +  \vec{\nabla}\cdot(\rho \vec{u} H_t )= \vec{\nabla}\cdot \left( \underline{\underline{\tau}}\cdot\vec{u}\right)+  \vec{\nabla}\cdot \left(k\vec{\nabla}T+\sum_{k=1}^{N_{sp}} \rho D h_k\vec{\nabla}Y_k  \right)  \nonumber\\
&\hspace{3.2 cm}-\sum_{k=1}^{N_{sp}} \dot{m}'''_k h_{f,k}^\circ -\vec{\nabla}\cdot \left(\rho Y_{C(s)} \vec{V}_{T} h_{C(s)}\right) - \kappa_P (\sigma T^4 - G)  \nonumber
\end{align}
\noindent
where, $k$ is the thermal conductivity, and $E = e + \vec{u}\cdot{\vec{u}}/2$ is the total sensible energy. $Y_k$ are the species mass fractions. $\underline{\underline{\tau}}~=\mu( \vec{\nabla}\vec{u} + (\vec{\nabla}\vec{u})^T-2/3~\underline{\underline{I}}(\vec{\nabla}\cdot\vec{u})) $
is the viscous stress tensor, $H_t = E + p/\rho $ is the total sensible enthalpy and $h_{f,k}^\circ$ is the heat of formation of the $k^{th}$ species calculated using NASA 7 coefficients \cite{NASA7}. An ideal gas equation of state is used to define the pressure, $P = \rho R_uT/MW$, with universal gas constant, $R_u$ (=8314.459 $J/kmol-K$) and mixture molecular weight, $MW = (\sum_{k=1}^{N_{sp}} Y_k/MW_k)^{-1}$, where $MW_k$ is the molecular weight of the $k^{th}$ species.

$\dot{m}'''_k$ is the mass consumption or production rate of the $k^{th}$ species from chemical reactions defined in compact notation as,
\begin{equation}
\dot{m}'''_k = MW_k\sum_{r=1}^{N_{rxn}} (\nu_{k,r}''-\nu_{k,r}')\dot{q}_r,
\end{equation}
where $\nu_{k,r}''$ and $\nu_{k,r}'$ are the stoichiometric coefficients on the products and reactants side of the $k^{th}$ species for reaction $r$. $\dot{q}_r$ is the reaction rate of progress variable for reaction $r$ defined as,
\begin{equation}
\dot{q}_r = k_{F,r}\Pi_{k=1}^{N_{sp}}\left[\frac{\rho Y_k}{MW_k}\right]^{\nu_{k,r}'} - k_{R,r}\Pi_{k=1}^{N_{sp}}\left[\frac{\rho Y_k}{MW_k}\right]^{\nu_{k,r}''},
\end{equation}
where $k_{F,r}$ and $k_{R,r}$ are the forward and reverse rate constants of reaction $r$ respectively. In this study, we used a detailed chemical kinetic mechanism for MMA oxidation developed by Bolshava \textit{et. al.} \cite{BCS21}, which consists of $N_{sp}=67$ species and $N_{rxn}=263$ reactions. The forward reaction rates for each reaction are modeled as either; modified Arrhenius reactions ($k_f = A T^b exp(-E_a/R_uT)$, three-body reactions, or Troe falloff reactions. Each of these forward reaction rates are explained in more detail in \cite{Turns00}. The constants used by each reaction are included as supplementary material to this paper in a Cantera \cite{cantera} yaml file format or can be found in CHEMKIN \cite{chemkin} formatted files in the original study \cite{BCS21}. $\vec{V}_{T}$ is the thermophoretic velocity associated with soot ($C(s)$) thermophoresis \cite{ABSD21},
\begin{equation}
\vec{V_T} = \frac{\mu}{2\rho T}\vec{\nabla}T.
\end{equation}
A six step semi-empirical soot model based on an acetylene precursor is adapted into the chemical mechanism that includes reactions for nucleation, agglomeration, surface growth and oxidation, which are given in full detail in \cite{ABSD21}.

The specific heat at constant pressure, and enthalpy are all mixture weighted, i.e, $c_p = \sum_{k=1}^{N_{rxn}}c_{p,k}Y_k$, where $c_{p,k}$ and $h_k$ are related through $h_k = h_{f,k}^o + \int_{298 K}^T c_{p,k}dT$, and determined using NASA 7 coefficients defined in the supplementary material. The viscosity, $\mu$, is calculated using a Sutherland viscosity transport model ($\mu_o = 1.716\times10^{-5}~kg/m-s,~T_o = 273K,~S_o=111 K$),
\begin{equation}
    \mu(T) = \mu_o \left(\frac{T}{T_o}\right)^{3/2}\frac{T_o+S_o}{T+S_o}.
\end{equation}
The species diffusivity coefficients are assumed to be equal for all gaseous species. The thermal conductivity, $k$, and gas species diffusivity, $D$, are determined from $\mu$, assuming a unity Lewis number and Prandtl number of 0.707. The soot species diffusivity is assumed to be mostly from thermophoresis, $D_{Y_{C(s)}}\vec{\nabla}Y_{C(s)} = Y_{C(s)}\vec{V_T}$.

The quantity $\kappa_P = \kappa_{P,gas} + \kappa_{P,soot}$ is the total Planck mean absorption coefficient including contributions from the gas and soot concentrations. $\sigma$ is the Stefan-Boltzmann constant and $G = \int_{4\pi} I d\Omega$ is the irradiation that is determined through solution of the radiative transfer equation (RTE) for gray gases \cite{Modest22}.  $\kappa_{P,gas}$ 
is computed using an emperical model provided by Zimmer \cite{Zimmer16} based on the concentrations of $CO_2,~H_2O,~CO$ and $CH_4$. $\kappa_{P,soot}$ is the Planck mean absorption coefficient of soot assuming a Rayleigh scattering limit \cite{Modest22}. 

The wall boundary conditions are simply treated as isothermal no-slip walls held at the ambient temperature of 300 K. The inlet and outlet boundary conditions are described using locally one-dimensional inviscid characteristics (LODI) \cite{POINSOT1992104}. The freestream velocity profile prescripted at the inlet is described using a $1/7^{th}$ power law for fully developed turbulent profile for pipe flow, $u_{in}(r) = V_{in}(1-2r/D_{pipe})^{1/7}$, where $V_{in} = 60~G_{ox}/49\rho_{ox}$ is a velocity factor defined by the oxidizer mass flux ($G_{ox}$) to obtain the correct mass flow rate, the oxidizer density is $\rho_{ox} = 1.283 \frac{kg}{m^3}$, and the inlet is $D_{pipe} = 0.0254m$.

The fuel boundary energy balance is used to couple the heat flux incident on the fuel surface from the gas phase to the mass of fuel vaporizing of the surface,
\begin{equation}
    \label{eqn:fuelBC}
    \dot{q}''_{c,g} + \dot{q}''_{r,Net} = \dot{m}''_f  l_v + \dot{q}''_{c,f}.
\end{equation}
where $\dot{q}''_{c,g} (= k \vec{\nabla} T \cdot \hat{n}_s)$ is the conductive heat flux from the gas phase normal to the fuel surface, $\dot{q}''_{c,f}$ is the conductive heat flux used in heating the fuel, and $\dot{q}''_{r,Net} (= \alpha_s\dot{q}''_{r,abs} - \epsilon_s\sigma T_f^4$) is the net radiation heat transfer at the fuel surface. $\dot{q}''_{r,abs}$ is the irradiant radiation heat flux onto the fuel surface, and $T_f$ is the fuel surface temperature. $\epsilon_s = 0.9$ and $\alpha_s=0.9$ are the fuel surface emissivity and absorptivity. $\dot{m}''_f = \rho_f \dot{r}$ is the local fuel mass flux and is related to the local regression rate by the solid fuel density $\rho_f = 1190 \frac{kg}{m^3}$. The current simulations assume the fuel is isothermal and already at the vaporization temperature of PMMA ($T_f = 653~K$) such that the is no heat loss due to heating the fuel ($\dot{q}_{c,f} = 0$). This leads to the following mass flux and regression rate relations depending only on the instantaneous gas phase solution,
\begin{equation}
    \label{eqn:RegressionRate}
    \dot{r}(x,y,t) = \frac{\dot{m}''_f(x,y,t)}{\rho_f} =\frac{k \vec{\nabla} T|_s(x,y,t) \cdot \hat{n}_s + \alpha_s\dot{q}''_{r,abs}(x,y,t) - \epsilon_s\sigma T_f^4}{\rho_f l_v} 
\end{equation}
In this isothermal fuel boundary model, the fuel temperature is held at $T_f = 653 K$ and the latent heat of sublimation at $l_v = 840890 J/kg$. The density, MMA species mass, and normal momentum boundary conditions have additional mass and momentum fluxes due to $\dot{m}''_f$. 

The system is solved using the DNS framework (ABLATE) which is available under a BSD-3-Clause License hosted on \href{https://github.com/UBCHREST/ablate}{GitHub.com/UBCHREST/ablate}. ABLATE utilizes a finite volume formulation on an unstructured grid employing the AUSM family of flux vector splitting schemes to solve the presented conservation equations. Time advancement is achieved using high-order Runge-Kutta methods.  Multi-block unstructured mesh domain decomposition is employed for high-efficiency parallel computation built upon PETSc’s DMPlex \cite{KnepleyLG15}.  Additional libraries include the open source library TChem \cite{kim:tchem:2021}, which is used to integrate the chemical kinetics required for the combustion modeling. 

The QoI the surrogates are trained on is a time-step weighted average of the regression rate at each location on the boundary $\mathcal{S}_B$. The QoI is selected as such to be able to later use the equivalent measurement from experimental data (regression rate between image snapshots) for calibration.
\begin{equation}
    \text{QoI:} \hspace{5mm} \tilde{\dot{r}}_{dt}(x,y) = \frac{\sum_{i=0}^{N} \dot{r}(x,y,t_i)\Delta t_i}{\sum_{i=0}^{N} \Delta t_i}    \hspace{10mm} \mathrm{for} \hspace{1mm} \mathrm{all} \hspace{1mm} (x, y)\in\mathcal{S}_B
\end{equation}
Where $\mathcal{S}_B$ includes the paired coordinates $x,y$ for the grid points on the boundary, $N$ is the number of simulation time steps, $t_i$ is the elapsed simulation time, and $\dot{r}(x,y,t)$ is the regression rate as computed during the simulation from Eqn. \ref{eqn:RegressionRate}.

\begin{figure}[h]
    \centering
    \includegraphics[width=.4\textwidth]{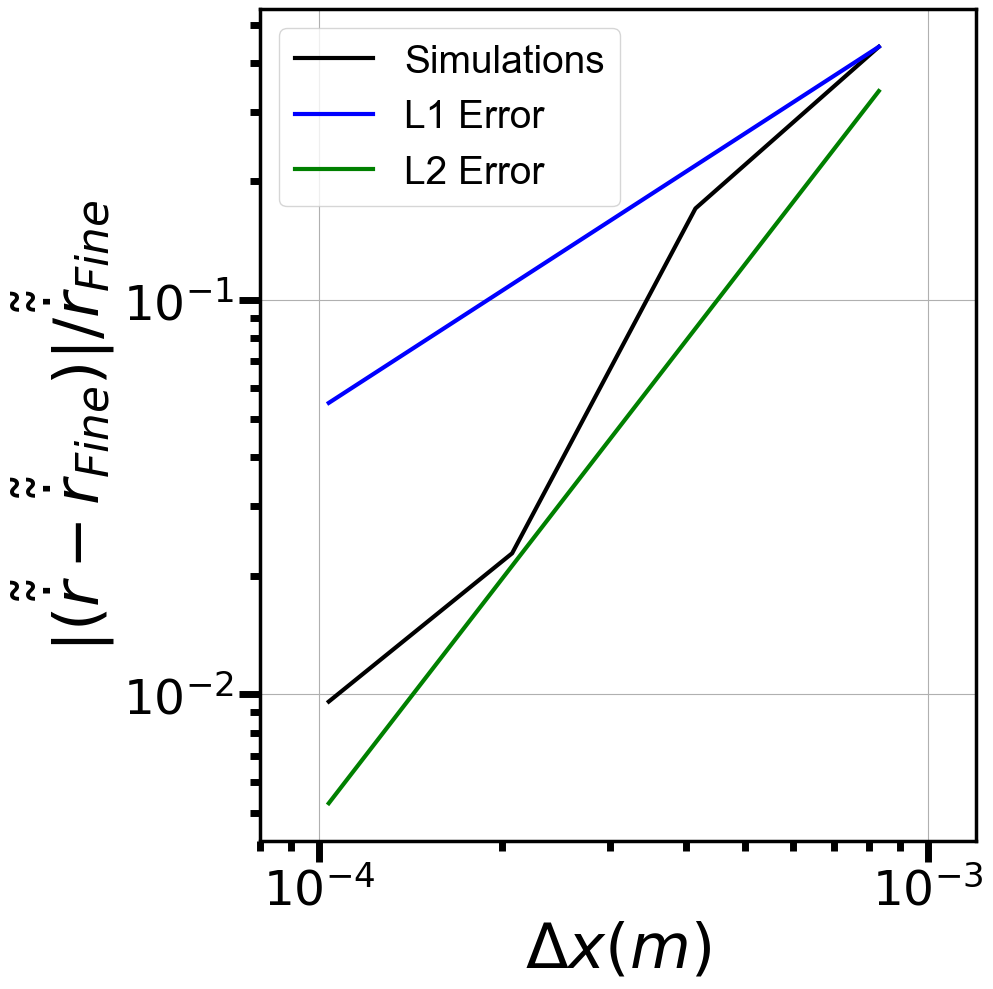}
    \caption{Relative error in $\tilde{\tilde{\dot{r}}}$ as a function of the grid step size compared to the finest mesh size case ($\Delta x$ = 0.05 $mm$) }
    \label{fig:MeshRefinement}
\end{figure}

A grid refinement study is performed to determine the size of the 2D mesh. Five different grid sizes are performed with average cell heights/widths of $\Delta x$ = 0.83, 0.41, 0.2, 0.1, and 0.05 $mm$. The regression rate QoI is averaged spatially and used as an error measure to test the mesh accuracy, $\tilde{\tilde{\dot{r}}} = \int_{\mathcal{S}_B}\tilde{\dot{r}}_{dt} d\ell/\int_{\mathcal{S}_B}d\ell$. The black line in Fig. \ref{fig:MeshRefinement} shows the relative error in $\tilde{\tilde{\dot{r}}}$  for the first four grid levels compared to the value from the finest mesh ($\Delta x$ = 0.05 $mm$), $\tilde{\tilde{\dot{r}}}_{Fine}$. Also shown in the blue and the green lines are the L1 and L2 expected error convergences respectively. The relative error typically follows a first order convergence due to using a first order gradient approximation for the conductive heat flux at the fuel surface. The mesh size of $0.1$ mm is used in this study since $\tilde{\tilde{\dot{r}}}$ is within 1\% of the fine case. With this mesh each simulation of the DNS resolves, on average, about 25-75ms of the reacting flow (1 to 3 flow through times) and takes around 24-48 hours depending mostly on the inlet flow velocity.

This section provides details of the surrogate models (Sections 3.1--3.2), the sampling ensemble strategy to train them (Section 3.3), and their implementation for facilitating forward UQ (Section 3.4) and Bayesian calibration (Section 3.5). All the data come from the DNS 2D slab burner reacting flow simulation described in the previous section.
\subsection{Gaussian Processes (GPs)}
GPs are widely used surrogates because they are fast, and therefore allow for UQ analyses of expensive computational models to be feasible.
GP assumes that the likelihood function of the QoI is a multivariate normal distribution (Eq. \ref{eq:Gasp}). 
\begin{equation}
    [y_{x,y}(x_1), y_{x,y}(x_2), ..., y_{x,y}(x_n)]^T \approx \mathcal{MN}([\mu(x_1), \mu(x_2), ..., \mu(x_n)]^T, \sigma^2 \boldsymbol{R})
    \label{eq:Gasp}
\end{equation}
Where $\mathcal{MN}(\cdot)$ is a multivariate normal distribution,$y_{x,y}(\cdot)$ is the real-valued QoI at a point location with coordinates $(x,y)$, $\boldsymbol{x^d} \in \mathcal{X}$ is the input vector to the simulation, $\mu (\cdot)$ is the mean function, $\sigma^2$ is the unknown variance, and $\boldsymbol{R}$ is the correlation matrix.
The mean function $\mu(\cdot)$ of a GP is typically modeled as a regression:
\begin{equation}
    \mu(\boldsymbol{x}) = \sum_{i = 1}^q h_i(\boldsymbol{x})\theta_i
    \label{eq:meanfunc}
\end{equation}
Where $h_i(\cdot)$ are the mean basis functions and $\theta_i$ the corresponding regression coefficient. For computational efficiency, the basis functions $h_i(\cdot)$ are common between points that belong to the same QoI and only the coefficients $\theta_i$ can vary. 
The elements of the correlation matrix $\boldsymbol{R}$ in (Eq. \ref{eq:Gasp}), are the values of a chosen correlation function between observation vectors $\boldsymbol{x_i}, \boldsymbol{x_j}$:
\begin{equation}
    \label{eq:corr}
    c(\boldsymbol{x_i}, \boldsymbol{x_j}) = \Pi_{k = 1}^d c_k(x_{ik}, x_{jk})
\end{equation}
Where $c_k$ is the output of the correlation function for the $k^{th}$ coordinate of the two input vectors. There are many options for correlation functions and here we use the rational quadratic kernel, which is created by adding squared exponential kernels across different length scales because it would capture some of the multiscale effects of the DNS.
Overall, the unknown parameters in the GP formulation include the regression coefficients $\theta$, the variance $\sigma^2$, and the range parameters from the kernel as a vector $\boldsymbol{\gamma}$. In this formulation, the marginal likelihood function after integrating out ($\theta, \sigma^2$) becomes:
\begin{equation}
    \label{eq:likelihood}
    \mathcal{L}(y^d|\boldsymbol{\gamma}) \propto |\boldsymbol{R}|^{-1/2} |\boldsymbol{h}^T(\boldsymbol{x}^d) \boldsymbol{R}^{-1}\boldsymbol{h}^T(\boldsymbol{x}^d)|^{-1/2} (S^2)^{-(\frac{n-q}{2})}
\end{equation}
Where $S^2 = (\boldsymbol{y^d})^T\boldsymbol{Q}\boldsymbol{y^d}$, $\boldsymbol{Q} = \boldsymbol{R}^{-1}\boldsymbol{P}$, and $\boldsymbol{P} = \boldsymbol{I_n}-\boldsymbol{h}(\boldsymbol{x^d}) [\boldsymbol{h}^T(\boldsymbol{x^d})\boldsymbol{R}^{-1}\boldsymbol{h}(\boldsymbol{x^d})]^{-1}\boldsymbol{h}^T(\boldsymbol{x^d})\boldsymbol{R}^{-1}$. $\boldsymbol{I_n}$ is the identity matrix of size $n=64$, which is equal to the number of simulation runs.
The range parameters $\gamma$ are then estimated by the modes of the marginal joint posterior distribution:
\begin{equation}
    \label{eq:mle}
    \hat{\boldsymbol{\gamma}} = argmax_{(\gamma_1,...,\gamma_p)}\mathcal{L}(y^d|\boldsymbol{\gamma})\pi(\gamma_1,...,\gamma_p)
\end{equation}
To train the GP, we use \textit{n} design points from the DNS ($\boldsymbol{x^d} = [\boldsymbol{x_1^d}, \boldsymbol{x_2^d}, ... ,\boldsymbol{x_{n}^d}]$) and the output is the QoI at these points $\tilde{\dot{r}}(\boldsymbol{x^d})$ for the parameter estimation described above. After training, the GP can be used for predicting the QoI at a specific position (x,y) on the slab burner boundary from a set of new input $\boldsymbol{x^*}$.
\subsection{Hierarchical Multiscale Surrogate (HMS)}
\begin{figure}
    \centering
\includegraphics[height=2in]{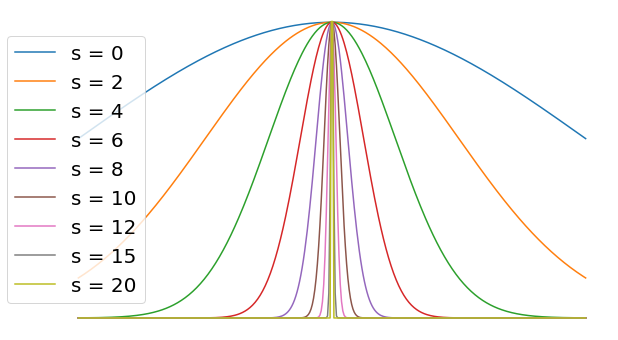}
    \caption{Behavior of the basis functions centered at the same point at increasing scales ($s$). Reprinted with permission from \cite{SHEKHAR2022115760}. Copyright 2022 Elsevier under Creative Commons License.}
    \label{fig:msbasis}
\end{figure}
Hierarchical and multiscale models typically have an identified hierarchy of approximations and make joint inference on the data by combining these model components in some rational fashion \cite{BERMANIS2013multiscale,FLOATER199665}. 
HMS introduces a scale parameter $s$ and defines a mapping between $s$ and the corresponding approximation space $\mathcal{H}_s$ in a hierarchy of spaces. The idea of exploiting the inherent correlation structure in the data at multiple levels follows directly from \cite{SHEKHAR2016887, SHEKHAR20171652}. The hierarchical models are able to analyze irregularly structured data at different resolutions (scales), rapidly minimizing the error in fitting using
  scale and data dependent basis functions $B^s$,   constructed by sampling suitable kernels at successively finer scales. Basis selection to approximate the observed data efficiently \cite{Shekhar2020b} uses a greedy scheme with forward selection and backward deletion phases that provides good performance  avoiding the performance bottleneck of other strategies \cite{SHEKHAR2022115760}. Fig. \ref{fig:msbasis} shows the behavior of the basis functions centered at the same point at increasing scales $s$.

  \begin{figure}[h!]
    \centering
    \includegraphics[width=\textwidth]{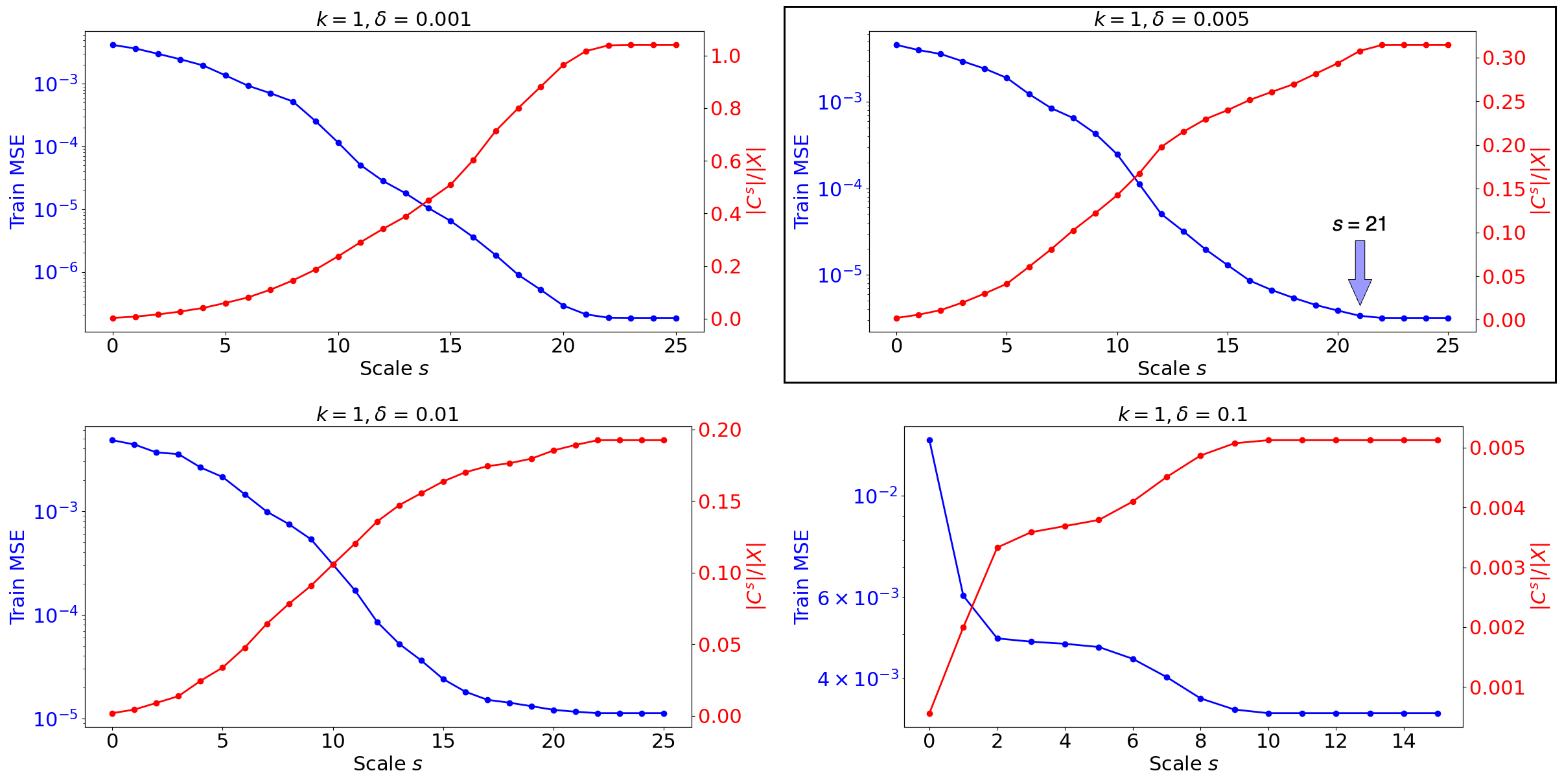}
    \caption{Hyperparameter tuning for the HMS surrogate. $\delta$ describes the upper bound tolerance for the current scale tolerance ($\epsilon_s < \delta$), $s$ is the number of scales, and $k$ refers to a training data fold. Based on these results, we select $s = 21, \delta = 0.005$ for the two hyperparameters.}
    \label{fig:HMStuning}
\end{figure}

\begin{algorithm}[!t]
	\caption{The Hierarchical Multiscale Approach \cite{SHEKHAR2022115760}}
	\label{alg:HMS}
\textbf{Input}: $D = (X,y)$, $\omega$, $\epsilon_0$ 
\textbf{Initialize}: $s = 0$, $B^s = [\cdot]$, $\Theta^s = [\cdot]$, $C^s = [\cdot]$, $Af^{s-1} = 0$, $t_s = y$\\

\textbf{While} $s \leq \omega$ \textbf{do:} \\
\textit{Step 1}. $B_s, \Theta_s, C_s, \mathcal{E} \leftarrow Forward\_selection(\epsilon_s,D,s,t_s)$ \label{step01}\\
\textit{Step 2}.  $B_s, \Theta_s, C_s \leftarrow Backward\_deletion(\epsilon_s,B_s,\Theta_s,C_s,\mathcal{E},t_s)$ \label{step02} \\
\textit{Step 3}. $B^s,\Theta^s, C^s \leftarrow Append(B_s,\Theta_s,C_s)$ \\
\textit{Step 4}. Current scale approximation: $f_s \leftarrow B_s \Theta_s$  \label{step04}\\
\textit{Step 5}. Current overall approximation: $Af^s \leftarrow Af^{s-1} + f_s$ \label{step05}\\
\textit{Step 6}. Update scale: $s \leftarrow s + 1$  \label{step06}\\
\textit{Step 7}. Update target for next scale: $t_s \leftarrow t_{s-1} - f_{s-1}$  \label{step07}\\
\textit{Step 8}. Update tolerance: $\epsilon_s$  \label{step08}\\
\textbf{End While} \\     

  where
  
$D$: Dataset (X,y), such that $X \in \mathbb{R}^{n \times d}$ and $y \in \mathbb{R}^n$,
 $s$: scale number,
  $\omega$: truncation scale,
 $\epsilon_0$: tolerance at scale 0,
 $Af^{s}$: final approximation till scale $s$,
  $t_s$: target at scale $s$,
  $B_s, \Theta_s, C_s$: basis set, corresponding weights, and sparse representation at scale $s$,
 $B^s, \Theta^s, C^s$: cumulative basis set, corresponding weights, and sparse representation till scale $s$
\end{algorithm}

Algorithm \ref{alg:HMS} summarizes the key steps.
A multiscale kernel (Eq. \ref{kernel1}) \cite{SHEKHAR2022115760}, which uniquely defines a customized associated approximation space (RKHS) is used.  {\it s}: scale of function, T: size of domain.
\begin{equation}
\label{kernel1}
K(x,y) = \sum_{s \in \mathcal{I}} \zeta_s \sum_{j \in r_s} \psi_j^s(x) \psi_j^s(y),\quad  \psi^s_j(z) = \exp \Bigg(-\frac{||z - z_j||^2}{\kappa_s} \Bigg), \quad \kappa_s = \frac{T}{2^s}
\end{equation}
  At each scale for a given target $t_s$,   the following optimization problem is solved
\[
\min_{\beta_s} ||t_s - K_s\beta_s||_2, \quad \text{subject to } ||\beta_s||_0 \leq p
\]
Here $K_s = [b_s^j]$ is the full set of n functions at scale $s, p$ are some positive constant and $||\beta_s||_0$ counts the number of non-zero values in $\beta_s$.
  Since, this is a non-convex problem,  an approximation of the problem corresponding to $||\cdot||_1$ norm is   obtained. The greedy forward selection starts with an empty basis set $B_s$, and recursively adds functions from the set $K_s$ which provide the highest reduction in the current residual. 
  This is continued until the tolerance  $\epsilon_s$ is   satisfied by the residual $r^t$ as 
$
\frac{|r^T b^j_s|}{||b_s^j||_2^2} < \epsilon_s
$.
Before training the HMS surrogate, we performed hyperparameter tuning to select the number of scales and tolerance bounds. The tuning process is about the trade-off between approximation error and the compression ratio of the data $|C^s|/|X|$. As more scales $s$ are added, the training error reduces, while the need for more data in the approximation increases. The selected number of scales and tolerance are selected such that we achieve an acceptable order of error magnitude ($ \l e^{-05})$, while also maintaining the compression ratio as low as possible for fast prediction. Based on our test runs (see Fig. \ref{fig:HMStuning} for results from some runs), we selected $s = 21$ scales and the tolerance bound as $\epsilon_s < \delta = 0.005$ because this combination achieves the lowest MSE with a reasonable compression ratio. $\delta = 0.001$ results in prohibitive compression ratios for a surrogate and $\delta = 0.01$ would be acceptable as well, but it is preferred to achieve a lower MSE for only 0.1 increase in the compression ratio. 

\subsection{Uncertain Inputs and Ensemble Strategy}
The relevant input quantities to ABLATE are summarized in Table \ref{tab:ablateinputs}. To create accurate surrogates, the simulation ensemble needs to be selected such that it covers the domain and codomain of the simulation as best as possible. Some of the input quantities are considered uncertain and therefore have to be sampled for the ensemble. For the ensemble runs, the geometry remains fixed because the simulation runs for a relatively small real time (on average 25-75ms). We do not expect the geometry change to be significant during this short time given that the maximum regression rate of PMMA experimentally is measured up to 0.15mm/s \cite{ZHANG2021337}, which justifies our assumption for the fixed geometry during a short simulation. We initially considered all possible reaction rate parameters of the Arrhenius expression (Eq. \ref{eq:ratecoeff}) and used local and global sensitivity analysis to select the reaction rate parameters to be included in the simulation ensembles, as described in Section 3.3.1. Only the selected reaction parameters are sampled for the ensembles, the remaining ones are fixed at expected values from literature. The inlet velocity factors are selected such that the ABLATE simulation corresponds to reasonable oxidizer flux $G$ of an equivalent slab burner experiment \cite{Georgalis_2023}, so we selected velocities that correspond to $G  {\in} [5, 20] \frac{kg}{m^2s}$. Of the fuel parameters $\theta_f$, the latent heat of sublimation $l_v$ is considered uncertain and sampled given an estimated range from expert input, but is later calibrated with available experimental data. To generate the 64 simulation ensemble, the uncertain input quantities were selected with Latin Hypercube Sampling (LHS), where each sample is the only one in each axis-aligned hyperplane containing it. The values for the uncertain inputs of each simulation of the ensemble are included in table format as supplemental material. 

\begin{table*}[!t]
\caption{Relevant input quantities for the ABLATE coupled flow/combustion solver.}
\label{tab:ablateinputs}
\centering
\renewcommand{\arraystretch}{1.2}
\begin{tabularx}{\textwidth}{p{6.5cm} p{1.5cm} p{1.6cm} p{5cm}}
\toprule
\toprule
\textbf{Quantity (units)} & \textbf{Symbol} & \textbf{Type} & \textbf{Description/Notes} \\
\midrule
Velocity factor (m/s) & $V_{in}$ & Uncertain & Velocity profile is assumed parabolic \\
Activation Energies (cal/mol)  & $E_*$ & Uncertain & Energy required to form the transition state for a given reaction \\
Temperature exponents (-) & $b_*$ & Uncertain & See Eq. \ref{eq:ratecoeff} \\
Pre-exponential factors (cm$^3$ mol$^{-1}$ cal $^{-1}$) & $A_*$ & Uncertain & Empirical constant associating temperature and rate coefficient \\
Geometry & $\mathcal{L}_{slab}$ & Fixed & - \\
Latent heat of sublimation (J/kg) & $l_v$ & Uncertain & The amount of heat energy required for PMMA to sublimate (fuel param.)\\
Sublimation temperature (K) & $T_f$ & Fixed & Temperature at which PMMA sublimates (fuel param.)\\
\bottomrule
\bottomrule
\end{tabularx}
\end{table*}

\subsubsection{Feature selection of Reaction Rate Parameters}
High fidelity models of fuel combustion typically involve a significant number of species and reactions. Despite the large dimensionality of the reaction rate parameter space, it has been observed that in many combustion systems, just a few parameters often have a more significant impact on the uncertainty of chemical model outputs compared to others \cite{gao2020uncertainty,zador2006local, ziehn2008global}. Based on the literature, we use sensitivity analyses and feature selection to systematically reduce the parameter space. The result is a reduced set of reaction rate parameters that we consider uncertain and include as features in the ensembles that train the surrogate models. The local sensitivity analysis (LSA) investigates the impact of small variations around a nominal parameter value on some reaction output (ignition delay). LSA employs a truncated Taylor expansion to approximate the sensitivity coefficient to a first-order level, which is advantageous due to its simplicity and fast computation speed, making it a valuable tool for gaining qualitative insights into the model behavior. Global uncertainty analysis (GSA) is a statistical approach and enables us to determine how uncertainties in ignition delay can be attributed to uncertainties in model parameters. The Sobol method \cite{sobol2007} is a variance-based global sensitivity analysis (VGSA) that ranks the importance of parameters by quantifying the extent to which the conditional variance caused by a parameter explains the variance in the model output. The Sobol method evaluates total sensitivity indices for each input parameter, providing estimates of the parameter's effect and its interactions with other parameters on the variation of ignition delay. Feature selection is a process to select a smaller set of features than those available and can be accomplished in many ways depending on the goal \cite{MIAO2016919}. By leveraging these methods to reduce the uncertain parameter space in chemical models we do the following, in order:
\begin{enumerate}
    \item Local sensitivity analysis to identify most influential reactions out of all reactions in the mechanism.
    \item Global sensitivity analysis on the parameters of the reactions identified in the previous step to find the most important rate constant parameters contributing to the chemical model output.
    \item Feature selection between the reaction rate parameters with the objective being to minimize the testing MSE of the HMS surrogate. The other slab burner covariates ($x, l_v, V_{in}$) all remain in the surrogate.
\end{enumerate} 

\begin{figure}[h!]
    \centering
    \begin{subfigure}[b]{0.48\textwidth}
        \centering
        \includegraphics[width=\textwidth]{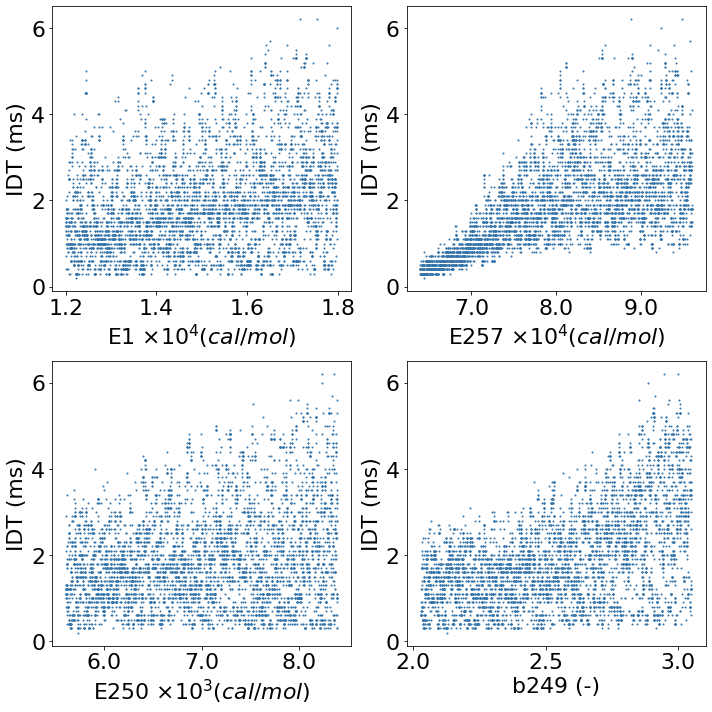}
    \end{subfigure}
    \hfill
    \begin{subfigure}[b]{0.48\textwidth}
        \centering
        \includegraphics[width=\textwidth]{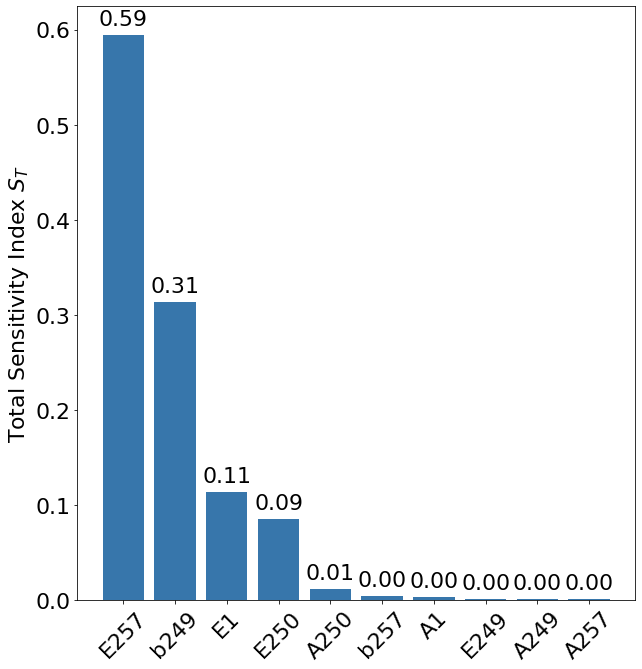}
    \end{subfigure}
    \caption{Left: Effect of the most important Arrhenius parameters on ignition delay for our MMA mechanism. Right: Total sensitivity indices of the Arrhenius parameters of the four influential reactions identified by local sensitivity analysis. The parameters ($E_{257}, b_{249}, E_1, E_{250}$) are considered most important and are therefore part of the uncertain simulation ensemble.}
    \label{fig:global_sensitivity}
\end{figure}

We have implemented this framework to identify sensitive reactions and parameters of a reduced MMA mechanism, consisting of 67 species and 263 reactions. The dependent variable the sensitivity analyses are completed with is the ignition delay time (IDT), due to its significance in combustion kinetics \cite{GURURAJAN2019478}. The first-order sensitivity index is the change of MMA mass fraction with respect to the rate constants ($k_i$) of the $i$th reaction, e.g., $\dfrac{\partial \ln(MMA)}{\partial \ln(k_i)}$, and is evaluated at the pressure of 1 atm and over a range of initial temperature of reactor between 600K and 2000K. 
Typically, forward rate constants ($k_i$) of the MMA mechanism are calculated via the 3-parameter Arrhenius expression: 
\begin{equation}
    k_i(T) = A_i T^{\beta_i} \exp\left(-\frac{E_i}{RT}\right),
    \label{eq:ratecoeff}
\end{equation}
where
$T$ is the temperature, and 
$E_i$, $A_i$, and $\beta_i$ are the
activation energy, pre-exponential factor, and temperature exponent coefficient of the $i$th reaction, respectively. By identifying the sensitive reactions  via local sensitivity analysis, it becomes computationally feasible to perform VSA on their associated associated rate parameters ($A_i, b_i, E_i$). 
To conduct VGSA, we sample all parameters from a uniform probability distribution function (PDF) and compute the total Sobol sensitivity index ${S_T}_i = \frac{\mathbb{E}[Var(IDT_i|k_{\sim i})]}{Var(IDT)}$. However, in the combustion literature,  it is known that there are multiple sources of correlation among these parameters \cite{prager2013uncertainty} {\cite{sutton2016effects}}.
For our VSA, we consider a 20\% uncertainty bound around the nominal values of the Arrhenius parameters associated with the most influential reactions and use TChem \cite{kim:tchem:2021} solvers for the sensitivity analyses. TChem computes the ignition delay time for a 0-dimensional constant volume homogeneous gas reactor. TChem defines IDT as the time when the build-up of $\mathrm{OH}$ radicals peaks.
From the LSA, the most important reactions are the following: \\
$$\mathrm{Reaction \hspace{1mm} \# 1: \hspace{1mm} H + O_2 \Leftrightarrow O + OH}$$
$$\mathrm{Reaction \hspace{1mm} \# 257: \hspace{1mm} MMA \Leftrightarrow T-C_3H_5 + CH_3OCO}$$
$$\mathrm{Reaction \hspace{1mm} \# 250: \hspace{1mm} MMA + CH_3 \Leftrightarrow MJ + CH_4}$$
$$\mathrm{Reaction \hspace{1mm} \# 249: \hspace{1mm} MMA + H \Leftrightarrow PJ + H_2}$$
Figure \ref{fig:global_sensitivity} illustrates the results from the VGSA. The total sensitivity index is evaluated using 6000 samples drawn from the parameter distributions using  {LHS}, using the Monte{-}Carlo  {(MC)} estimator proposed by \cite{saltelli2002making} and implemented in the parallel object-oriented library DAKOTA \cite{dakota2020}.
The figure shows that the activation energy in reaction 257 exhibits the most impact on ignition delay. This particular MMA reaction corresponds to chain branching highly reactive species, which play a crucial role in sustaining the combustion process by breaking down and producing radicals. Other important parameters include the activation energies of reaction 1 and 250 as well as the temperature exponent for reaction 249. Furthermore, we observed the rest of the parameters ($A_{250}, b_{257}, A_1, E_{249}, A_{249}, A_{257}$) to have  a negligible effect on the {IDT}. Consequently, these parameters can be considered deterministic when producing simulation samples to train a surrogate model, as their uncertainties do not significantly impact the overall variability in the {IDT}. 

\begin{figure}[h!]
    \centering
    \includegraphics[scale = 0.4]{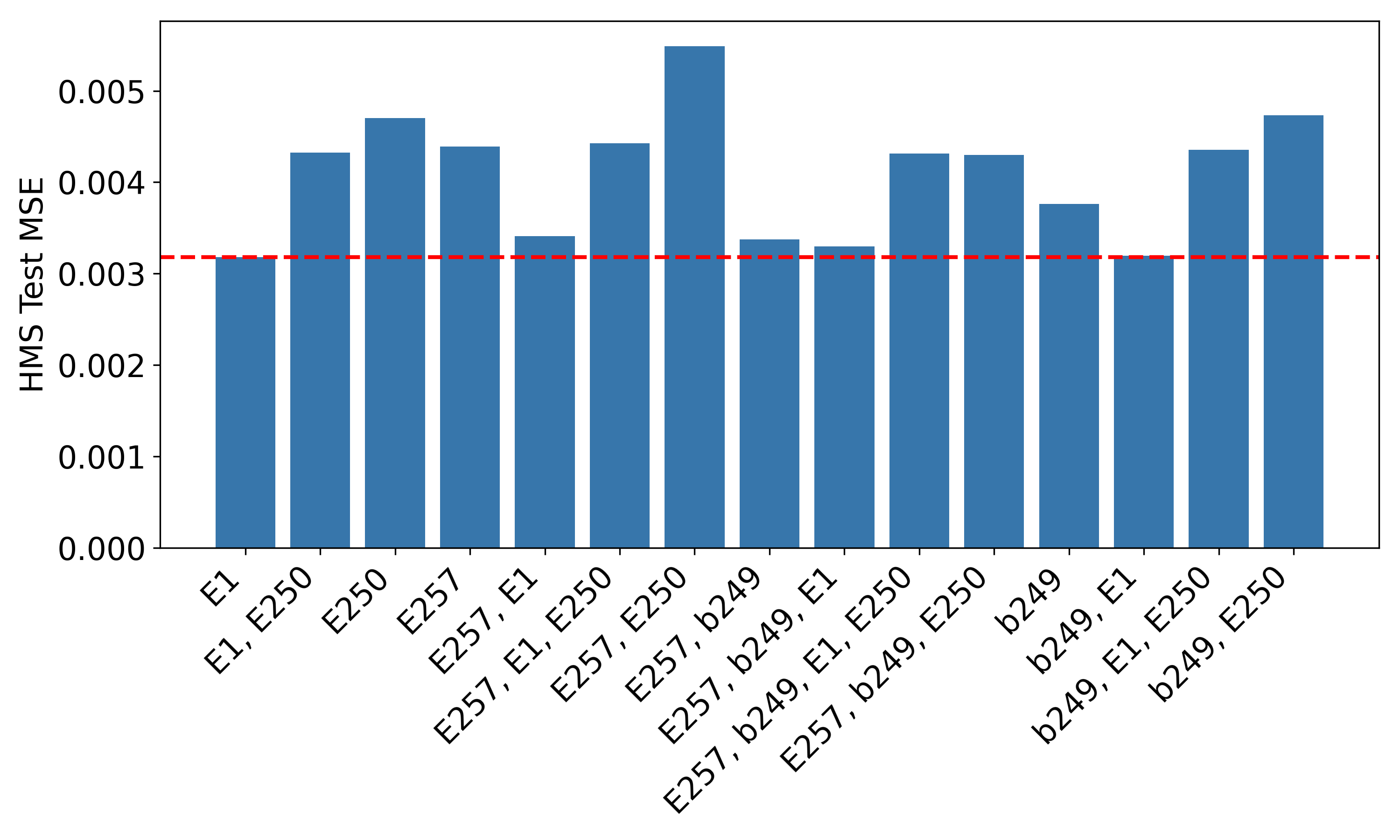}
    \caption{Surrogate feature selection between reaction rate parameters. The dashed red line indicates the minimum MSE. $E_1$ and $(E_1, b_{249})$ achieve similar results. The test MSE is added over all testing folds. Based on these results, we select $E1$ as the only reaction rate to be included in the surrogate.}
    \label{fig:featureselect}
\end{figure}

To complete the feature selection, the testing MSE of the HMS surrogate is evaluated across different testing folds covering the entire dataset. Each testing fold is predicted with different models arising from the different combinations of reaction rate parameters and all the other slab burner predictor features ($x, l_v, V_{in}$). The testing MSE evaluated here is an addition of the error at each fold. The results of the feature selection are shown in Fig. \ref{fig:featureselect}. Based on our analysis, we only include $E_1$ as the only reaction rate feature in the surrogate models. While it may be unexpected that $E_1$ alone produces the best surrogate model even if it is not the parameter with the highest sensitivity index, the sensitivity analysis (which is conducted with ignition delay as the dependent variable) is only a part of the selection process, whereas directly testing the surrogate model against the slab burner QoI (the regression rate) is ultimately the deciding factor, as regression rate is the quantity we want to predict.

\subsection{Forward Uncertainty Propagation}
After the development of the surrogates, they can be used to propagate forward the uncertainty in the inputs $\boldsymbol{x} = [x, E_1, l_v, V_{in}]$ to the QoI $\tilde{\dot{r}}$. The goal is to approximate the probability distribution of the regression rate and its statistics, which can be achieved via  {MC} sampling methods \cite{https://doi.org/10.1002/wics.1539} (Alg. \ref{alg:ForwardUQ}). Forward UQ is possible given that the surrogates are fast to sample. To know when the MC sampling has converged, we monitor the mean and variance of the posterior distribution for the regression rate and stop when those its statistics difference are within a tolerance of $\epsilon_* = e^{-5}$. For the inputs, the velocity factor is assumed uniform throughout the parameter range. The fuel and chemistry parameters $l_v, E_1$, are assumed normal with the mean being the nominal default DNS value and a variance that results in the distribution covering values $\pm 25\%$ from that nominal value. Given that the regression rate is highly dependent on the location of interest as known from our previous work with experiments, we consider the results in different regions of the slab burner. Experimentally, the front part ($x \leq 30$mm) is expected to achieve the highest regression due to the ramp effects and ignition location, the middle part ($30 < x \leq 70$ mm) mostly achieves lower regression due to steady state combustion, and the rear part ($x > 70$mm) also has higher regression due to re-circulation effects. In each case, we uniformly sample the possible boundary locations.

\begin{algorithm}[!t]
	\caption{Forward UQ: Estimation of the distribution of the regression rate using the surrogate}
	\label{alg:ForwardUQ}
Given a portion of the slab (front, middle, or rear) with locations $(x,y) \in\mathcal{S}_B$ \\
Assume prior distributions for: \\
$ V_{in} \sim \mathcal{U}(V_{min} = 3.917 \frac{m}{s} ,V_{max} = 15.398 \frac{m}{s} )$ \\
$ E_1 \sim \mathcal{N}(15610,107.1 ) (cal/mol)$ \\
$l_v \sim \mathcal{N}(840890, 70074.16) \frac{J}{kg}$ \\
\textbf{Notes:} $V_{min}$ corresponds to an oxidizer flux of $G_{min} = 5 \frac{kg}{m^s}$ and $V_{max}$ to an oxidizer flux of $G_{max} = 20 \frac{kg}{m^2s}$. The standard deviation for the priors of $l_v$ and $E_1$ are selected such that the resulting distribution has sufficient range to cover $\pm$25$\%$ from the estimated nominal values of $lv= 840890 \frac{J}{kg}$ and $E_1 = 15610$ $cal/mol$.\\

\textbf{While} $\|\sigma_{\boldsymbol{y^*}}^{2, new} - \sigma_{\boldsymbol{y^*}}^{2,old}\| \geq \epsilon_{\sigma^2}$ \textbf{AND}  $\|\mu_{\boldsymbol{y^*}}^{new} - \mu_{\boldsymbol{y^*}}^{old}\| \geq \epsilon_{\mu}$ \textbf{do}: \\
\textit{Step 1}. Draw a sample from each of the input quantities to create the new design vector $\boldsymbol{x^*} = [x, E^*_{1}, l_v^*, V_{in}^*]$ \\
\textit{Step 2}.  Get the new prediction for the regression rate from the surrogate: $\boldsymbol{y^*} = \tilde{\dot{r}}(\boldsymbol{x^*})$  \\
\textit{Step 3}. Append corresponding probability distribution $p({\tilde{\dot{r}}}) =\mathrel{+} \boldsymbol{y^*} $ \\
\textit{Step 4}. Update statistics $\sigma_{\boldsymbol{y^*}}^{2,new}, \sigma_{\boldsymbol{y^*}}^{2,old}, \mu_{\boldsymbol{y^*}}^{new}, \mu_{\boldsymbol{y^*}}^{old}$ \\
\textbf{End While} \\            
\end{algorithm}

\subsection{Model Calibration under Uncertainty}
Finally, we demonstrate the Bayesian calibration of two key parameters: the latent heat of sublimation ($l_v$)--a fuel parameter in the DNS model—and the activation energy for reaction 1 ($E_1$)--a parameter in the chemical reaction mechanism. Both parameters are calibrated using regression rate measurements obtained from slab burner experiments. The calibration of physics-based models necessitates addressing uncertainties arising from incomplete and noisy experimental data, as well as the model's limitations in capturing complex physical phenomena. Traditional deterministic inverse methods are insufficient for quantifying these uncertainties. Therefore, we adopt Bayesian approaches for statistical inverse analysis, as outlined in \cite{oden2017predictive, tan2022predictive, liang2023bayesian}, which provide a probabilistic framework for UQ in model calibration.
The solution to the Bayesian calibration problem is the probability distribution functions (PDFs) of the parameters ($\boldsymbol{x}_{calib} = {l_v, E_1}$) given the experimental data $\mathbf{D}$, such that
\begin{equation}
    \pi_{post}(\boldsymbol{x}_{calib}|\mathbf{D}) = 
    \frac{\pi_{like}(\mathbf{D} | \boldsymbol{x}_{calib})  \pi_{prior}(\boldsymbol{x}_{calib})}
    {\pi_{evid}(\mathbf{D})},
    \label{eq:bayes}
\end{equation}
where the prior $\pi_{prior}(\boldsymbol{x}_{calib})$ represents initial knowledge about parameters before observing the data,
likelihood $\pi_{like}(\mathbf{D} | \boldsymbol{x}_{calib})$ is derived from a noise model representing data and model uncertainty, and 
evidence PDF $\pi_{evid}(\mathbf{D})$ is a normalization factor ensuring the posterior $\pi_{post}(\boldsymbol{x}_{calib}|\mathbf{D})$ is a PDF.

\begin{figure}[h!]
    \centering
    \includegraphics[scale = 0.35]{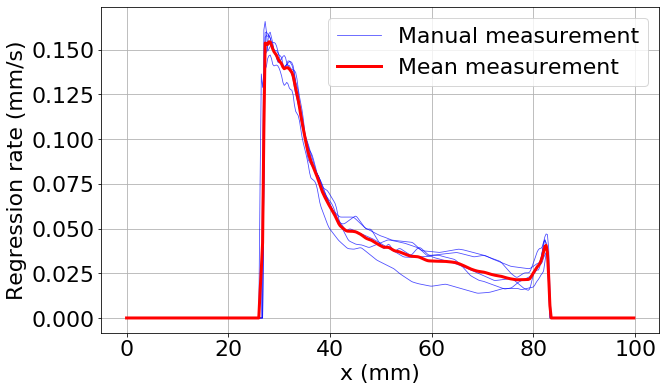}
    \caption{Available regression rate measurements from an experiment with oxidizer flux $G=8.08 \frac{kg}{m^2s} (\equiv V_{in} = 4.587 m/s)$.}
    \label{fig:expdata_calib}
\end{figure}

The primary computational challenges in the Bayesian calibration of coupled flow and combustion DNS models like ABLATE include the high computational cost of each forward run and the extensive parameter space, which makes sampling-based Bayesian algorithms like Markov Chain Monte Carlo (MCMC) computationally prohibitive. To address these challenges, we use the HMS surrogate model that substantially reduces the computational cost of mapping inputs to regression rate,
i.e., $\tilde{\dot{r}} = \mathbf{Y}(\boldsymbol{x}_{calib})$ within the Bayesian solution. Assuming an additive noise model and Gaussian random variables for the data noise with zero mean, we postulate the following form for the likelihood:
\begin{equation}
    \ln \left(\pi_{like}(\mathbf{D} | \boldsymbol{x}_{calib})\right)
    \propto
    \left[ \mathbf{Y}(\boldsymbol{x}_{calib}) - \mathbf{D} \right]^T 
    \boldsymbol{\Gamma}^{-1}
    \left[ \mathbf{Y}(\boldsymbol{x}_{calib}) - \mathbf{D} \right],
    \label{eq:noise}
\end{equation}
where $\boldsymbol{\Gamma}$ is a diagonal covariance matrix encapsulating the noise at each data point.
%
The process to measure the regression rate from image data is outlined in \cite{Georgalis_2023} and includes capturing snapshots of the PMMA solid fuel profile as it regresses during the burn using high-speed cameras. The resulting profiles are then either manually traced and segmented or are segmented using machine learning \cite{SURINA2022160}. The data used for this work were all manually traced. By knowing the height of the fuel profile over time and at each location, the regression rate on the boundary can be estimated by averaging the height difference over the observed experimental time of the snapshots. The available experiment data are shown in Fig. \ref{fig:expdata_calib}, taken for an experiment with oxidizer flux $G=8.08 \frac{kg}{m^2s}$.

We note that we have taken many steps to guarantee the closest possible equivalence between the regression rate from the surrogate and the experimental data, although there are a few differences that need to be disclosed. The simulation data (which the surrogate is trained with) are estimates of the regression from the heat flux balance at the slab burner boundary and are computed over short simulation times (typically a few tens of milliseconds). To the contrary, experimental regression rate is purely based on the fuel profiles, carries uncertainty due to the manual tracing, and the snapshots used to compute the regression rate estimates span over a few seconds due to waiting for the fuel specimen temperature to get to steady-state. With such data available, the steps we have taken to guarantee the closest possible equivalence between the two sets of data are the following: a) remove time-dependence on the regression rate by time-averaging either the profile difference (experiments) or the direct regression rate estimate (simulation), b) have the simulation use an isothermal boundary condition to best approximate the steady-state conditions of the experiment data collection, and c) estimate the variance in the experimental data used in the Bayesian calibration from 5 repetitions of the manual tracing of the slab profiles (see Fig. \ref{fig:expdata_calib}).

\section{Results and Discussion}
This section provides the results after implementation of the methods introduced in Section 3. The section starts with a comparison via cross validation between the two surrogates GP and HMS. Then, we use HMS for forward UQ to estimate the probability distribution of the regression rate from the uncertain inputs in 3 regions of the slab burner: front, middle, and rear. Lastly, we demonstrate Bayesian inference of the fuel parameter latent heat of sublimation ($l_v$) and the activation energy of reaction 1 ($E_1$) with experimental slab burner regression rate data.
\subsection{GP vs. HMS comparison}
To compare the two surrogate models, we first evaluate their training and testing accuracy on the available dataset via 5-fold cross validation \cite{CV}. We split the dataset of the 64 DNS simulations in 5 folds, 4 of which include the 13/64 simulations as part of the testing set and the remaining 51/64 simulations as training. The last fold includes 12/64 simulations as testing and the remaining 52/64 simulations as training. When a simulation is picked to be part of a training/testing set, all the boundary locations are included with it, we did not sample based on location. With the cross-validation approach, the goal is to evaluate the resulting surrogate models on the entire available dataset and particularly their ability to generalize and predict correctly out-of-sample, i.e., from data they have not been trained on. 

\begin{figure}[htbp!]
    \centering
    \includegraphics[width = \textwidth]{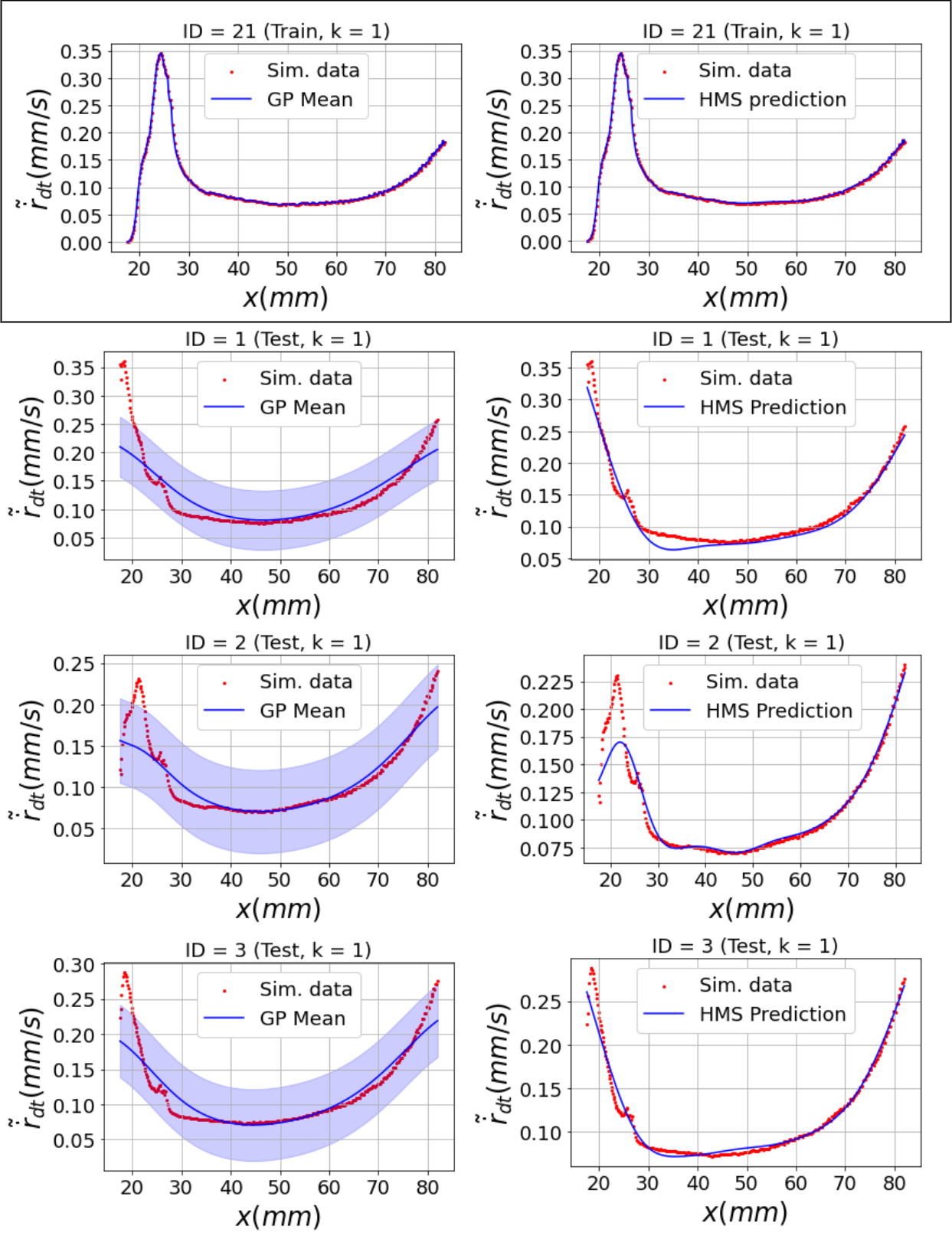}
    \caption{Comparison of GP surrogate and HMS for select simulation IDs belonging to the first fold of the cross validation. Both surrogates predict the training set perfectly (shown in bounded box). When the regression rate includes nonlinear effects, the HMS mean follows the trend much closer than the GP and is overall more accurate.}
    \label{fig:HMSvGP_1}
\end{figure}

\begin{figure}[htbp!]
    \centering
    \includegraphics[width = \textwidth]{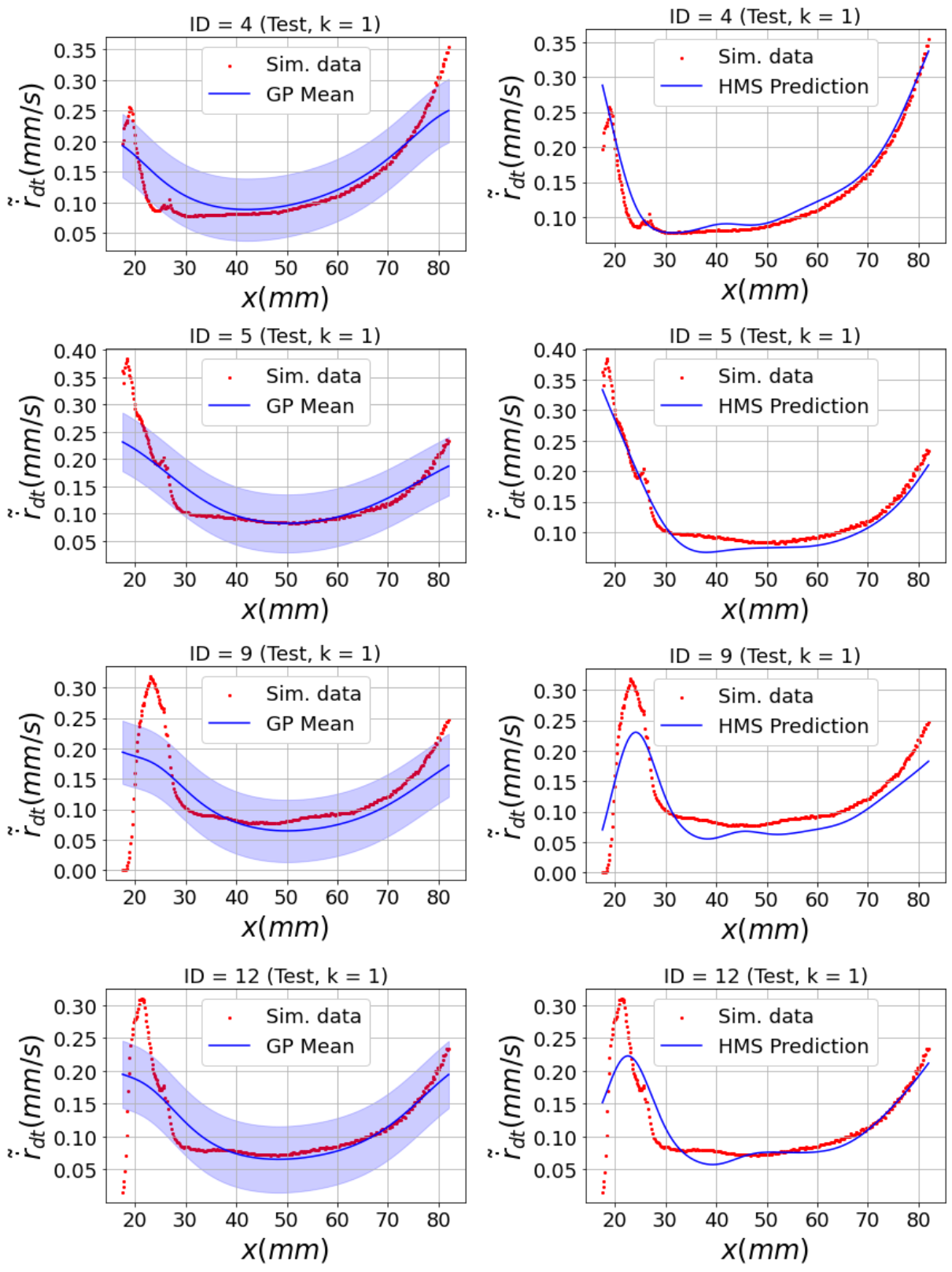}
    \caption{Further comparison of GP surrogate and HMS for select simulation IDs belonging to the first fold of the cross validation. When the regression rate includes nonlinear effects, the HMS mean follows the trend much closer than the GP and is overall more accurate.}
    \label{fig:HMSvGP_2}
\end{figure}

Figs. \ref{fig:HMSvGP_1}, \ref{fig:HMSvGP_2}  show some of the results from the first fold for both GP and HMS. Both surrogates predict the training set perfectly as expected. One obvious difference between the two models is that GPs inherently predict with a standard deviation which can be translated to a $95 \%$ confidence interval, unlike HMS. For this particular problem, due to the multiscale nature of the data, the GP confidence intervals are extremely wide in all predictions. The wide bounds indicate low confidence in the model predictions, but can be useful in practice to identify an accurate bounded range for the regression rate since they cover most of the data. Comparing the GP mean with HMS, we conclude that both models appear to have similar performance out of sample for the portion of the slab after the initial non-linear peak of the regression rate. We also conclude that the GP is unable to predict the non-linear regression rate in the front of the slab for all IDs. The HMS predicts the non-linear part of the slab better as evident in IDs $= 1, 3, 4, 5$. In some cases (see IDs $= 2, 9, 12$) the non-linear part at the front is not fully captured by the HMS, but the model form is able to follow the response, but miss the magnitude peak. We show some more comparisons between the two models from the other cross-validation folds in supplemental material. Overall, we conclude that the HMS due to its ability to capture multiscale effects is a better surrogate for accurate prediction as it is able to follow the regression rate response better out-of-sample. The information from the confidence interval of a GP is also useful in practice, as it provides a wide but valid range for the regression rate response. 

\begin{figure}[htbp!]
    \centering
    \includegraphics[scale = 0.7]{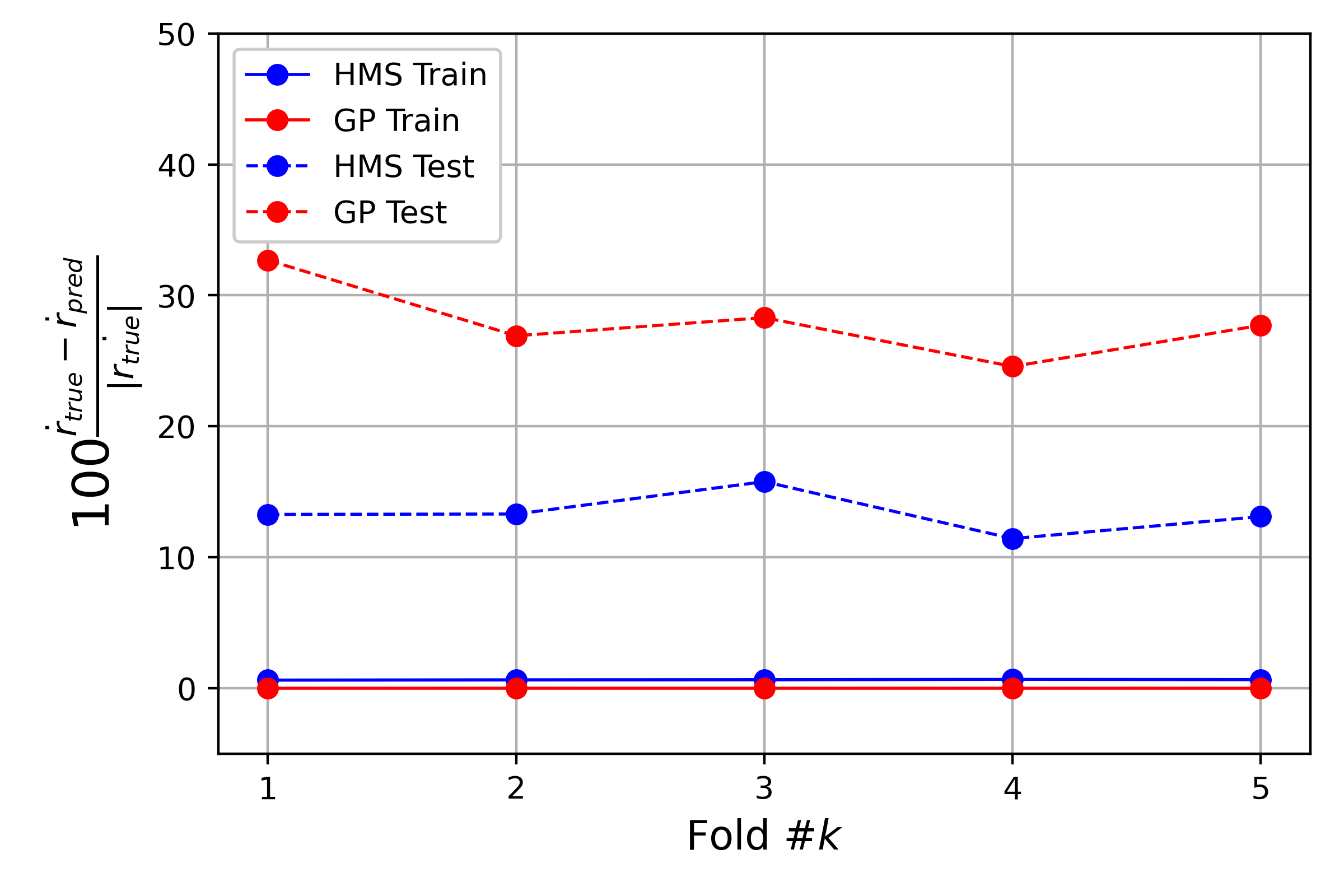}
    \caption{Performance of GP and HMS across all data folds of the cross validation. The GP, on average, is expected to achieve an error of $1.00 \times 10^{-8} \%$ in the training set and $28.03 \%$ when tested out of sample. The HMS, on average, is expected to achieve an error of $0.63 \%$ in the training set and $13.37 \%$ when tested out of sample.}
    \label{fig:CV}
\end{figure}

Fig. \ref{fig:CV} shows the training and testing performance of the two models across all the data folds, measured as a normalized percentage error from the true value of the prediction. The GP, on average, is expected to achieve an error of $1.00 \times 10^{-8} \%$ in the training set and $28.03 \%$ when tested out of sample. The HMS, on average, is expected to achieve an error of $0.63 \%$ in the training set and $13.37 \%$ when tested out of sample. The HMS is robust in prediction throughout the entire dataset. Overall, our conclusion is that HMS is better than the GP for usage in UQ or otherwise for prediction in this multiscale problem. The fact that HMS is able to achieve an error $<15 \%$ when predicting multiscale boundary quantities just from a few far field inputs is significant, particularly considering that it takes a few $ms$ to run compared to 24 hours for the DNS. Information from the GP is useful, if needing to bound the quantity within some range.

\subsection{Forward UQ with HMS} 
\begin{figure}[htbp!]
    \centering
    \includegraphics[scale = 0.22]{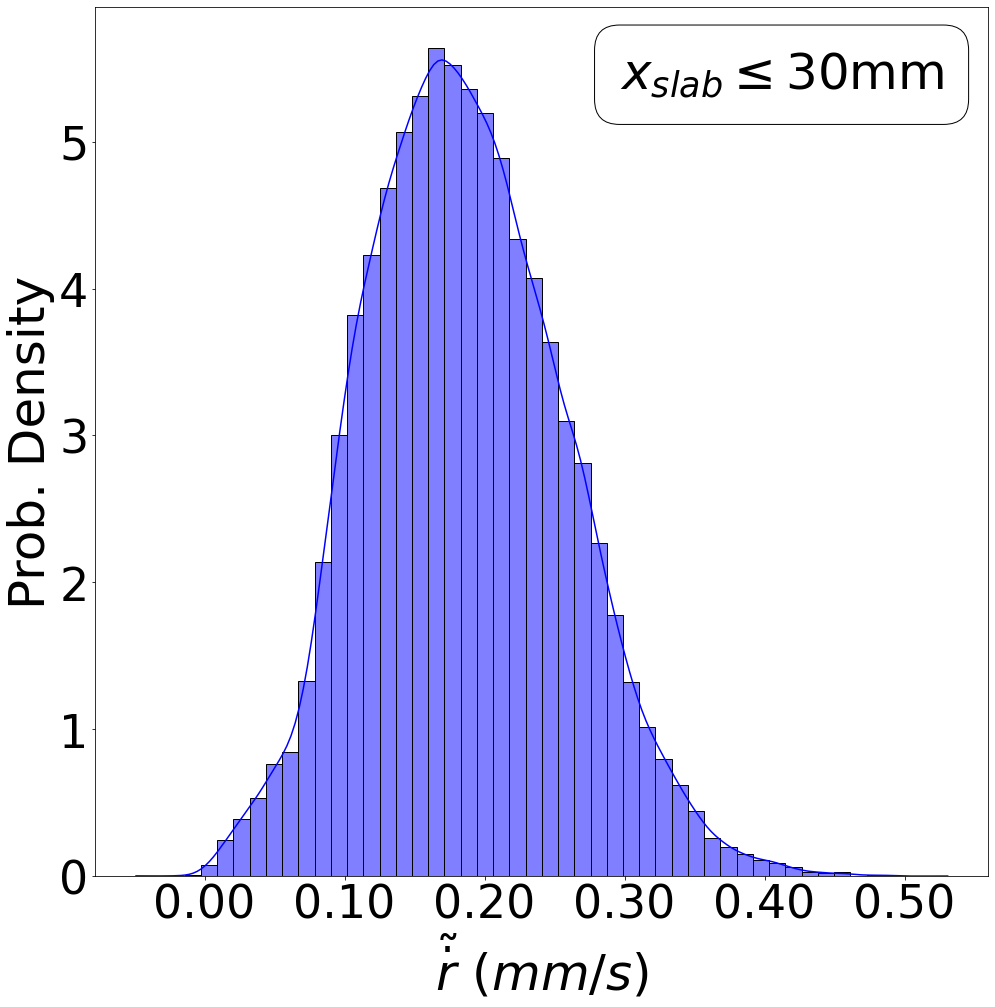}
    \caption{Forward propagation of uncertainty to the regression rate from uncertain inputs at the front of the slab burner corresponding to $G = [5, 20] \frac{kg}{m^2s}$.}
    \label{fig:ForwardUQ1}
\end{figure}

\begin{figure}[htbp!]
    \centering
    \includegraphics[scale = 0.4]{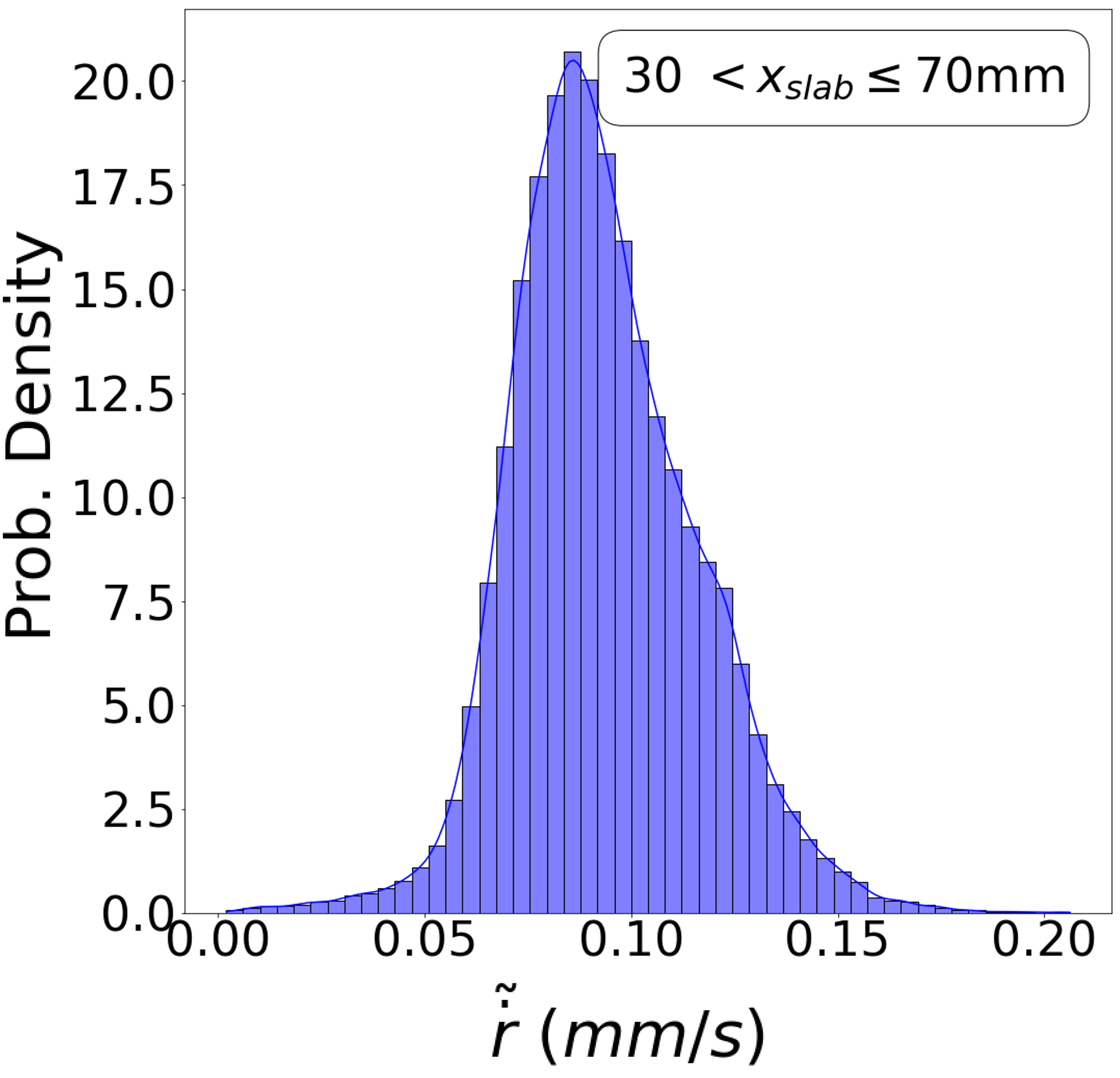}
    \caption{Forward propagation of uncertainty to the regression rate from uncertain inputs at the middle of the slab burner corresponding to $G = [5, 20] \frac{kg}{m^2s}$.}
    \label{fig:ForwardUQ2}
\end{figure}

\begin{figure}[htbp!]
    \centering
    \includegraphics[scale = 0.3]{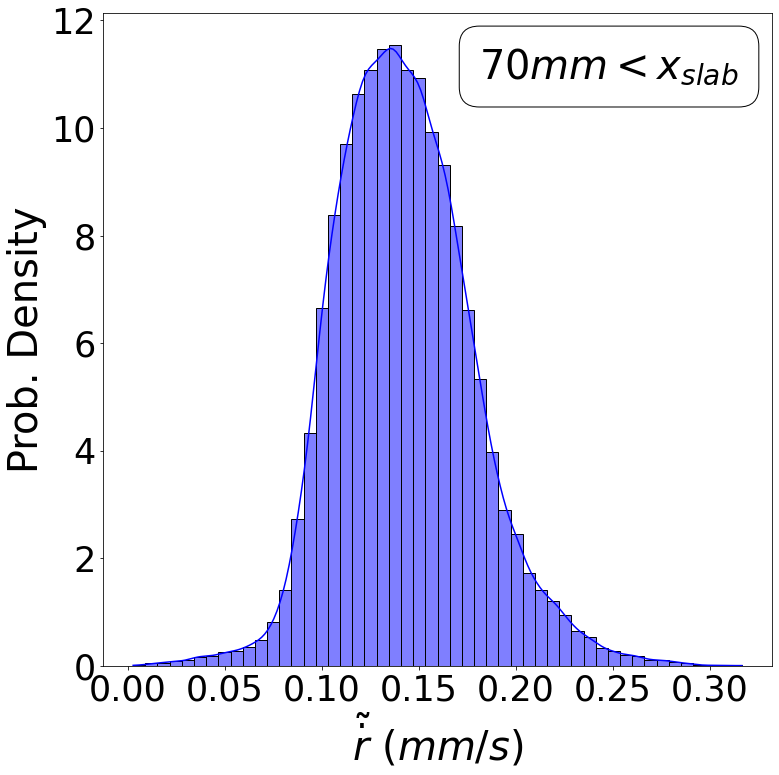}
    \caption{Forward propagation of uncertainty to the regression rate from uncertain inputs at the rear of the slab burner corresponding to $G = [5, 20] \frac{kg}{m^2s}$.}
    \label{fig:ForwardUQ3}
\end{figure}

In this section, we demonstrate the use of the HMS for propagating uncertainty from the inputs to the regression rate as shown in Alg. \ref{alg:ForwardUQ}. Approximating the distribution of the regression rate from the uncertain input space is useful for decision making and in this case to be able to bound the expected performance of the slab burner in terms of regression rate. We consider 3 areas of study for the slab burner since the regression rate is highly dependent on the location. Experimentally, the front part ($x \leq 30$mm) of the slab is expected to achieve the highest regression due to the ramp effects, the middle part ($30 < x \leq 70$ mm) mostly achieves lower regression due to steady state combustion, and the rear part ($x > 70$mm) also has higher regression due to re-circulation effects from the oxidizer flow. In each forward propagation, we select boundary locations uniformly that adhere to the restrictions of the location. The velocity factor $V_{in}$ is also uniformly sampled and its range corresponds to an oxidizer flux of $G_{min} = 5 \frac{kg}{m^s}$ and $G_{max} = 20 \frac{kg}{m^2s}$, which are appropriate for this slab burner. The remaining two parameters are the latent heat of sublimation $l_v$ and the activation energy for reaction 1 ($E_1$) which are assumed normal with the mean being the nominal default value of the DNS and a variance that corresponds to a distribution that covers $\pm 25 \%$ above or below the nominal value. The forward propagation uses traditional Monte-Carlo sampling and the results for the distributions of the regression rate as shown in Figs. \ref{fig:ForwardUQ1}, \ref{fig:ForwardUQ2}, \ref{fig:ForwardUQ3}. In the front of the slab, the estimate is a mean regression rate $\mu_{f} = 0.185$ mm/s, in the middle of the slab $\mu_{m} = 0.093$ mm/s, and in the rear $\mu_{b} = 0.142$ mm/s. All distributions appear to have a central tendency and follow the orders of magnitude we observe in the experiments at the corresponding locations.

\subsection{Bayesian model calibration using regression rate measurement data}
\begin{figure}[htbp!]
    \centering
    \includegraphics[scale = 0.35]{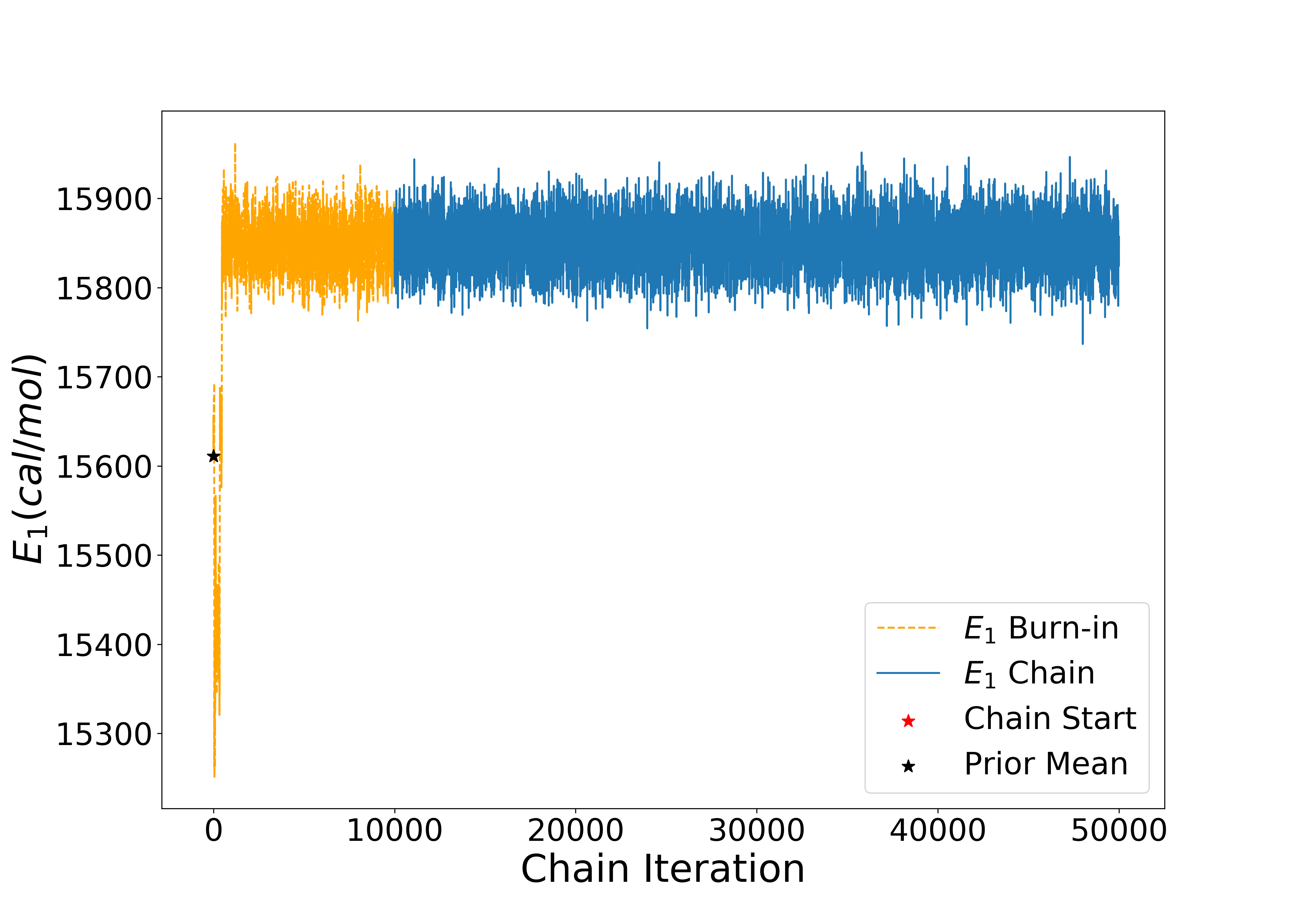}
    \caption{MCMC chain for $E_1$ starting at prior mean value.}
    \label{fig:E1_chain}
\end{figure}

\begin{figure}[htbp!]
    \centering
    \includegraphics[scale = 0.35]{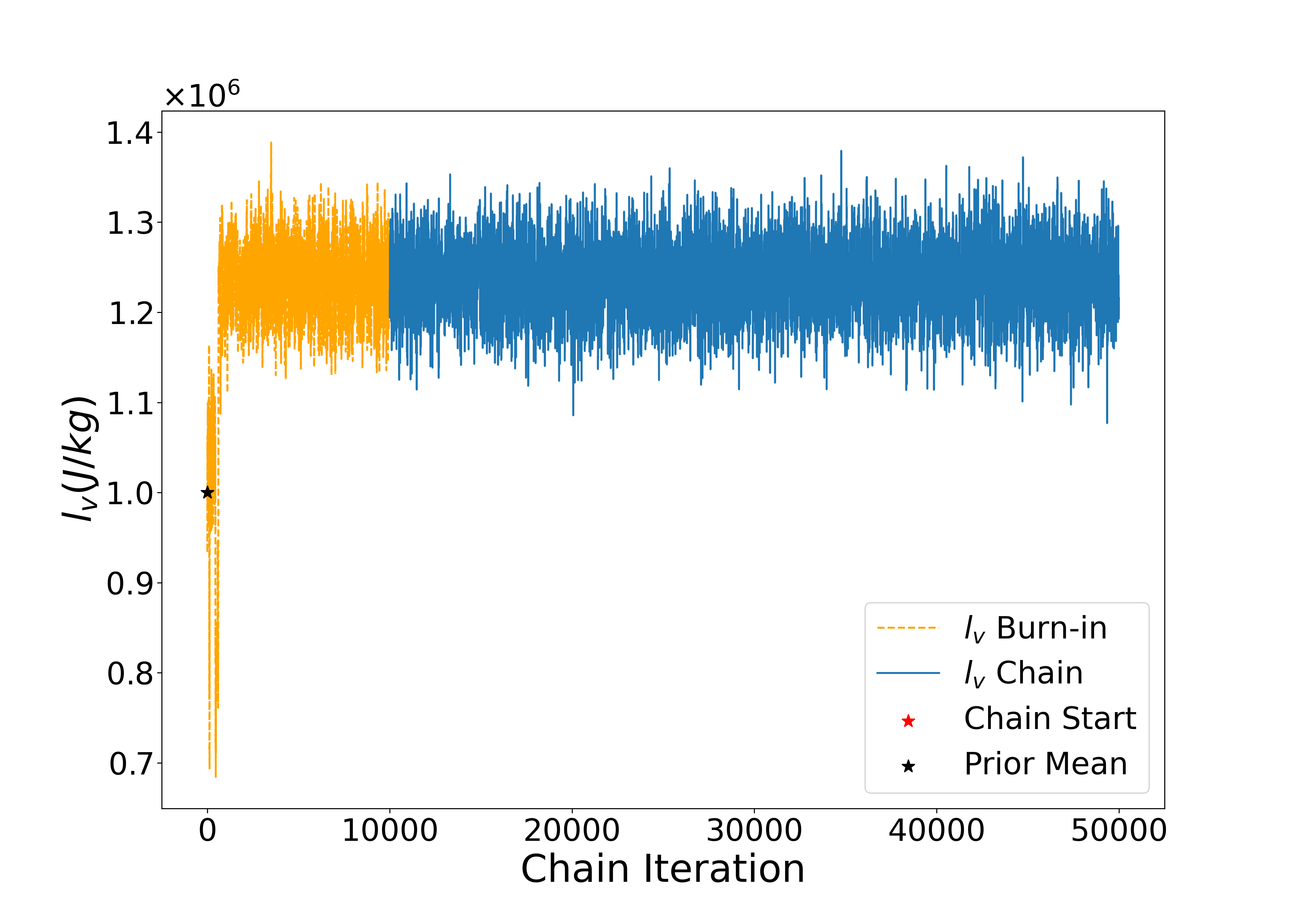}
    \caption{MCMC chain for $l_v$ starting at prior mean value.}
    \label{fig:lv_chain}
\end{figure}

\begin{figure}[htbp!]
    \centering
    \includegraphics[scale = 0.33]{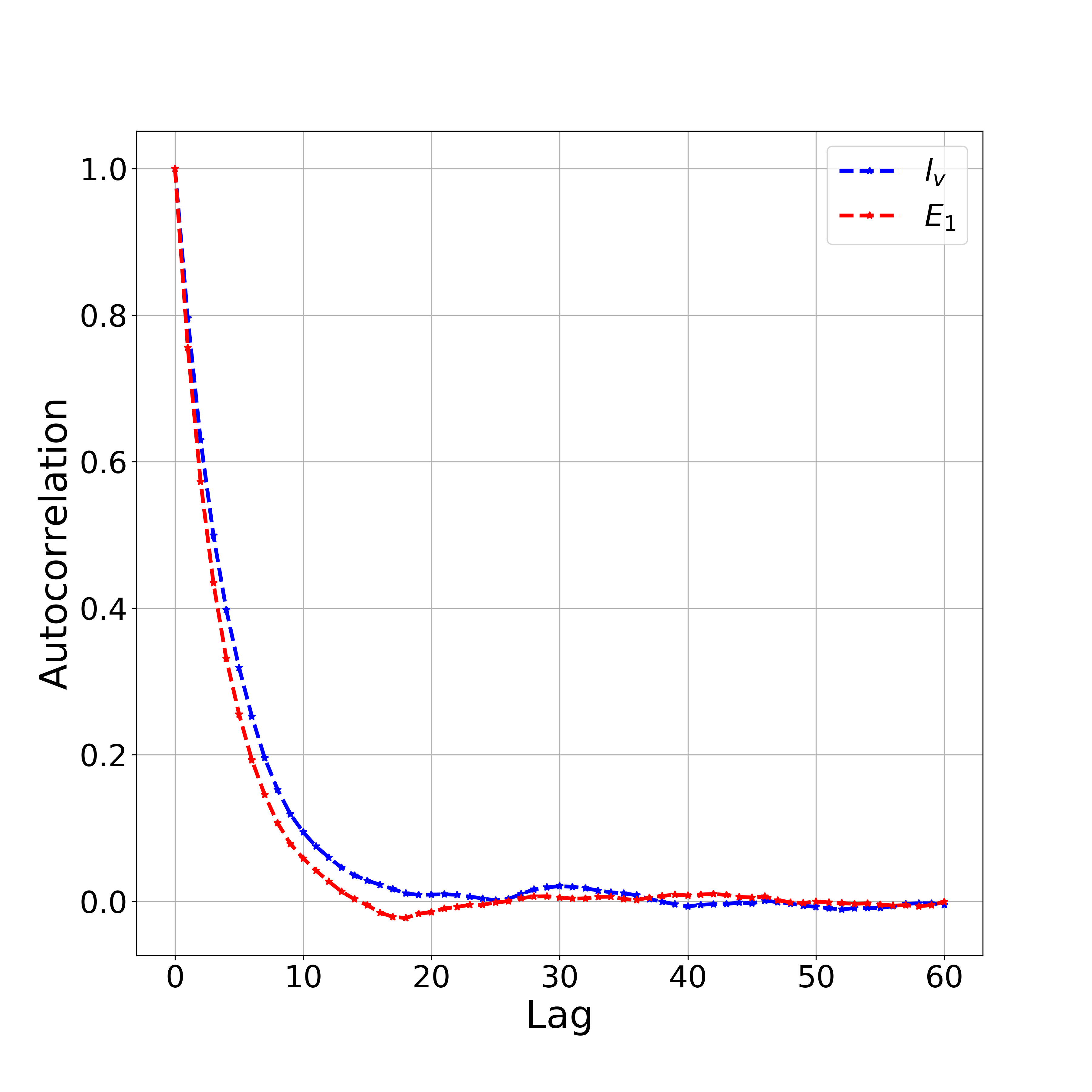}
    \caption{Autocorrelation plot of $lv, E_1$ MCMC chains.}
    \label{fig:BCauto}
\end{figure}

In this section, we use the HMS surrogate model in conjunction with available experimental data (Fig. \ref{fig:expdata_calib}) to calibrate two key physical parameters of the DNS model: the latent heat of sublimation $l_v$ and the activation energy $E_1$. As described in Section 2.5, we apply Bayesian inference to calibrate these parameters while quantifying the associated uncertainty in model parameters. The chosen prior distributions for the parameters $ E_1 \sim \mathcal{N}(15610, 107.1) (cal/mol)$ and
$l_v \sim \mathcal{N}(1e^6, 119369) \frac{J}{kg}$, which are wide enough to capture all possible values from literature \cite{SARI2014217}. To solve the Bayesian calibration problem, we employ the Random Walk Metropolis algorithm \cite{Haario1999}, a Markov Chain Monte Carlo (MCMC) method used to draw samples from the posterior distribution. To thoroughly explore the posterior distributions, we run 3 independent chains of 50,000 iterations each, starting from different initial points within the prior: both parameters at the minimum, mean, and maximum of their feasible domain. The first 20\% of each chain is discarded as ``burn-in,’’ which is standard practice to allow the chain to converge towards the true posterior before sampling begins.
Figs. \ref{fig:E1_chain}, \ref{fig:lv_chain} present trace plots for one representative MCMC chain initialized with both parameters at the mean of the prior range. Additionally, the autocorrelation plot (Fig. \ref{fig:BCauto}) is provided as a diagnostic tool to assess the independence of the samples. As shown, samples spaced 60 iterations apart exhibit minimal correlation, indicating that the MCMC chain is progressing efficiently and producing low-correlated, independent samples, and that the chain has appropriately explored the posterior distribution.

\begin{figure}[htbp!]
    \centering
        \includegraphics[scale = 0.35]{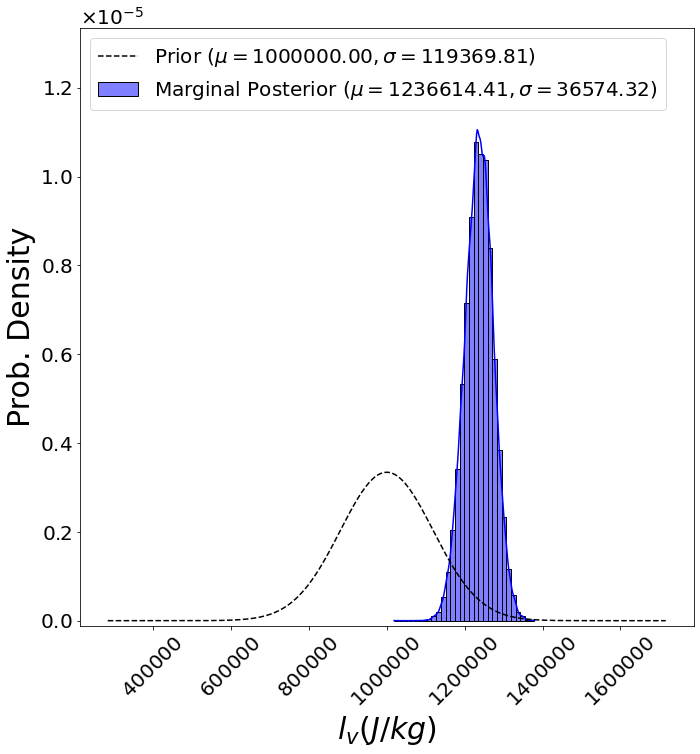}
        \caption{Final marginal posterior distribution of $l_v$ after Bayesian calibration using HMS and experimental data. The calibration shows that $l_v$ should be much higher value (increased by $30\%+$) than the default value of the parameter in the DNS.}
        \label{fig:finalposteriorlv}
    \hfill
\end{figure}

\begin{figure}[htbp!]
        \centering
        \includegraphics[scale=0.35]{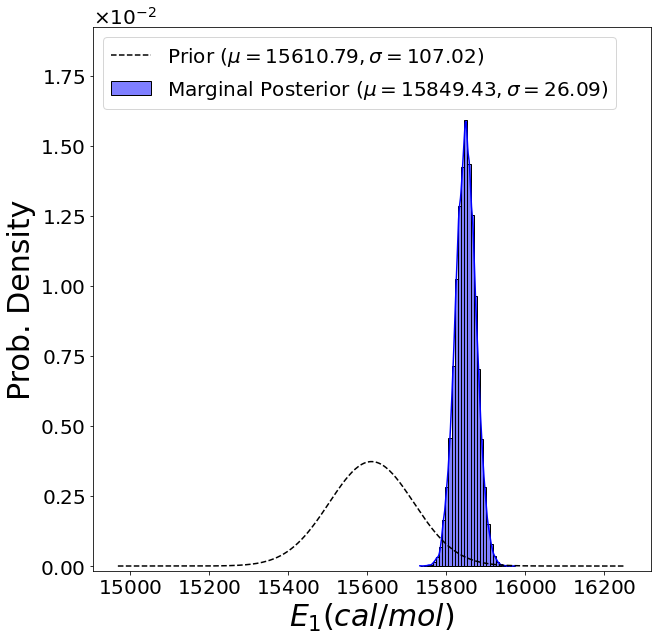} 
        \caption{Final marginal posterior distribution of $E_1$ after Bayesian calibration using HMS and experimental data. The calibration shows that $E_1$ should be slightly higher (~$1-2 \%$) than the default value of the parameter in the DNS.}
        \label{fig:finalposteriorb}
\end{figure}

\begin{figure}[htbp!]
    \centering
    \includegraphics[scale = 0.35]{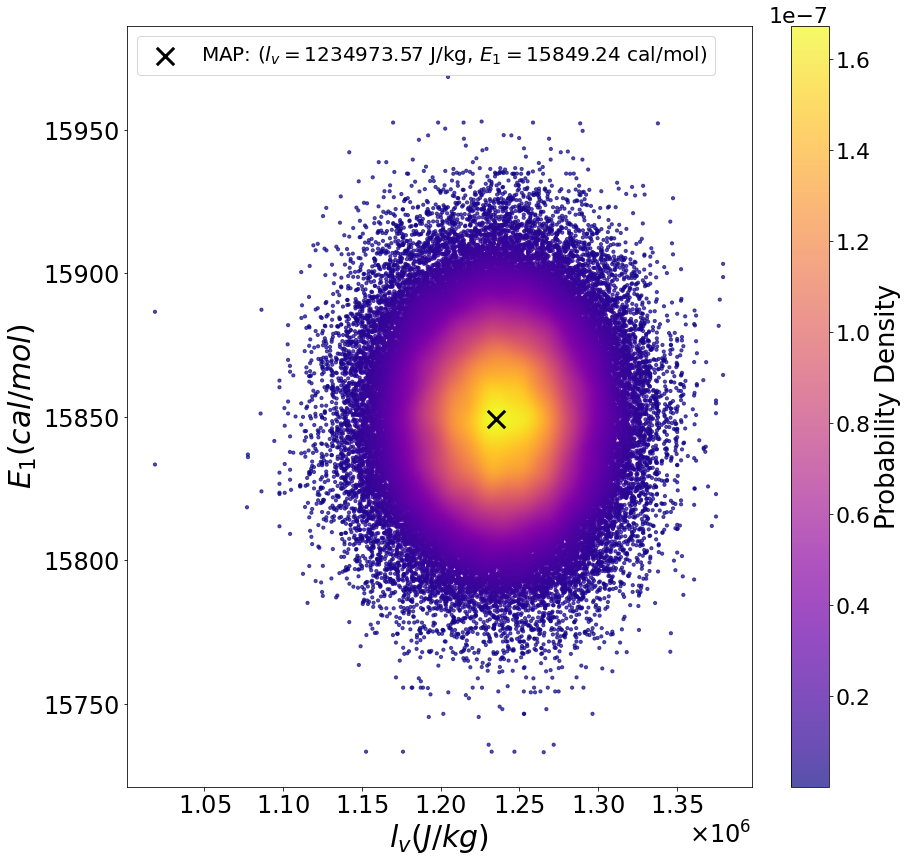}
    \caption{Final joint posterior distribution of $E_1$ and $l_v$ after Bayesian calibration using HMS and experimental data. The maximum aposteriori estimate (MAP) is at ($l_{v,MAP} = 1234973.57 \frac{J}{kg}, E_{1,MAP} = 15847.24 \frac{cal}{mol}$)}.
    \label{fig:2dMAP}
\end{figure}

To arrive at the final marginal posterior estimates (see Figs. \ref{fig:finalposteriorlv}, \ref{fig:finalposteriorb}), we  combine all 120000 samples from all 3 chains after burn-in. A main conclusion from the results is that the default values for both parameters are smaller than they should be to best allow the surrogate to approach the experimental results. Instead, the values should be higher based on the maximum aposteriori estimate (MAP) of the joint distribution ($l_{v,MAP} = 1234973.57 \frac{J}{kg}, E_{1,MAP} = 15847.24 \frac{cal}{mol}$) as shown in Fig. \ref{fig:2dMAP}.

\section{Conclusion and Future Work}
The main goal of this paper was to systematically address challenges related to all aspects of performing a complete UQ analysis of a complicated physics-based simulation like a 2D slab burner DNS. We provided insights related to the development of an acceptable surrogate model, the propagation of uncertainty of the DNS inputs to the QoI, and the calibration of DNS parameters based on observations of experimental data as well as interpreting these results for further DNS development. The DNS we used in this work was ABLATE (Ablative Boundary Layers At The Exascale), an in-house DNS solver developed for simulating a multiphase 2D slab burner setup. 

First, we generated an ensemble of 64 simulations that was informed by a series of sensitivity analyses to identify the most important reaction rate parameters. The simulation data was then used to develop two surrogate models: a Gaussian Process (GP) and a hyperparameter tuned Hierarchical Multiscale Surrogate (HMS). Our expectation was that the HMS would outperform the GP because of the multiscale effects of combustion. We tested both models under cross-validation for their ability to predict out of sample in parameter spaces they have not been trained on. We found that, overall, HMS is superior to GP for prediction. The GP, on average, is expected to achieve an error of $1.00 e^{-8} \%$ in the training set and $28.03 \%$ when tested out of sample. The HMS, on average, is expected to achieve an error of $0.63 \%$ in the training set and $13.37 \%$ when tested out of sample. Our conclusion is that HMS is superior to GP for usage in UQ or otherwise for this problem, due to its ability to better capture multiscale effects. HMS also demonstrated robust performance across the dataset. The fact that HMS is able to achieve an error $<15 \%$ when predicting multiscale boundary quantities just from far field inputs is significant, particulary considering it takes a few $ms$ to run compared to 24 hours for the DNS. 

Second, we used the HMS in a forward UQ process to propagate uncertainty from the inputs to the QoI which is the regression rate. We assumed default priors for the input parameters (normal or uniform distributions as appropriate). At the front of the slab, the estimate is a mean regression rate $\mu_{f} = 0.185$ mm/s, in the middle of the slab $\mu_{m} = 0.093$ mm/s, and in the rear $\mu_{b} = 0.142$ mm/s. All distributions appear to have a central tendency and follow the orders of magnitude we observe in the experiments at the corresponding locations. 

Last, we used the Bayesian approach to calibrate two of the parameter inputs, the latent heat of sublimation $l_v$ and the activation energy $E_1$ with available experimental data. We found the maximum aposteriori estimate (MAP) for these two parameters to be at ($l_{v,MAP} = 1234973.57 \frac{J}{kg}, E_{1,MAP} = 15847.24 \frac{cal}{mol}$). These MAP values are higher than the default used in the DNS (more than $30\%$ higher for $l_v$ and $1-2\%$ higher for $E_1$). Overall, the conclusion is that there is not one set of ($l_v, E_{1}$) that result in correctly predicting the regression rate in all locations of the slab simultaneously.

In the future we aim to expand on the work shown here through various efforts. Considering the combustion of higher alkanes ($C_nH_{2n+2}$) such as paraffin  wax in the same slab burner setup can lead to better understanding of hybrid rocket systems. Fuels like paraffin are also of interest due to higher regression rate compared to other polymers, and therefore UQ of the regression rate from the DNS and validation with equivalent experiments are also relevant. The need to investigate surrogates that are able to produce accurate estimates of time-varying field QoIs is also of interest, specifically when these surrogates need to be coupled as parts of sequential frameworks. Thus, future work can explore applications of our framework where a correlated, joint PDF of parameters estimated by fitting experimental or first-principles density-functional theory (DFT) calculations data using Bayesian inference techniques \cite{prager2013uncertainty, nagy2011uncertainty}  {\cite{medford2014assessing}}. Lastly, a more in depth calibration effort with more experimental data would help understand better how the surrogates perform across a wide range of conditions as well as identify the conditions that surrogates are appropriate for.

\section*{Acknowledgements} This work was supported by the United States Department of Energy’s (DoE) National Nuclear Security Administration (NNSA) under the Predictive Science Academic Alliance Program III (PSAAP III) at the University at Buffalo, under contract number DE-NA0003961.

The authors would also like to acknowledge the Tufts University High Performance Compute Cluster which was used for running the uncertainty analyses. We thank Kolos Retfalvi, Jasper Stedman, Alex Post, and Joseph Cygan from the University at Buffalo for their help with acquiring and tracing the experimental data used for the Bayesian calibration.

\bibliographystyle{unsrt}  
\bibliography{references}  

\begin{thebibliography}{10}

\bibitem{MAZZETTI2016286}
Alessandro Mazzetti, Laura Merotto, and Giordano Pinarello.
\newblock Paraffin-based hybrid rocket engines applications: A review and a market perspective.
\newblock {\em Acta Astronautica}, 126:286--297, 2016.

\bibitem{Karabeyoglu98}
M~Karabeyoglu, D~Altman, and D~Bershader.
\newblock Transient combustion in hybrid rockets.
\newblock In {\em 31st Joint Propulsion Conference and Exhibit}, AIAA-95-2691, San Diego, CA, July 1995.

\bibitem{OKNINSKI2021260}
Adam Okninski, Wioleta Kopacz, Damian Kaniewski, and Kamil Sobczak.
\newblock Hybrid rocket propulsion technology for space transportation revisited - propellant solutions and challenges.
\newblock {\em FirePhysChem}, 1(4):260--271, 2021.

\bibitem{CARRICKFUELS}
P.~Carrick and C.~Larson.
\newblock Lab scale test and evaluation of cryogenic solid hybrid rocket fuels.
\newblock {\em 31st Joint Propulsion Conference and Exhibit}, AIAA-95-2948, San Diego, CA, USA, July 1995.

\bibitem{Georgalis_2023}
Georgios Georgalis, Kolos Retfalvi, Paul~E. Desjardin, and Abani Patra.
\newblock Combined data and deep learning model uncertainties: An application to the measurement of solid fuel regression rate.
\newblock {\em International Journal for Uncertainty Quantification}, 13(5):23--40, 2023.

\bibitem{SURINA2022160}
Gabriel Surina, Georgios Georgalis, Siddhant~S. Aphale, Abani Patra, and Paul~E. DesJardin.
\newblock Measurement of hybrid rocket solid fuel regression rate for a slab burner using deep learning.
\newblock {\em Acta Astronautica}, 190:160--175, 2022.

\bibitem{Kobaldrheological}
M.~Kobald, E.~Toson, H.~Ciezki, S.~Schlechtriem, S.~di~Betta, M.~Coppola, and L.~DeLuca.
\newblock Rheological, optical, and ballistic investigations of paraffin-based fuels for hybrid rocket propulsion using a two-dimensional slab-burner.
\newblock {\em Progress in Propulsion Physics}, 8:263--282, 2016.

\bibitem{GLASER2023186}
C.~Glaser, R.~Gelain, A.E.M. Bertoldi, Q.~Levard, J.~Hijlkema, J.-Y. Lestrade, P.~Hendrick, and J.~Anthoine.
\newblock Experimental regression rate profiles of stepped fuel grains in hybrid rocket engines.
\newblock {\em Acta Astronautica}, 204:186--198, 2023.

\bibitem{Karabeyoglu_Exp}
Arif Karabeyoglu, Greg Zilliac, Brian~J. Cantwell, Shane DeZilwa, and Paul Castellucci.
\newblock Scale-up tests of high regression rate paraffin-based hybrid rocket fuels.
\newblock {\em Journal of Propulsion and Power}, 20(6):1037--1045, 2004.

\bibitem{KARABEYOGLU1}
M.A. Karabeyoglu, D.~Altman, and B.J. Cantwell.
\newblock Combustion of liquefying hybrid propellants: Part 1, general theory.
\newblock {\em Journal of Propulsion and Power}, 18(3):610--620, 2012.

\bibitem{Echekki_2009}
Tarek Echekki.
\newblock Multiscale methods in turbulent combustion: Strategies and computational challenges.
\newblock {\em Computational Science $\&$ Discovery}, 2(1):013001, 2009.

\bibitem{PETERS20091}
N.~Peters.
\newblock Multiscale combustion and turbulence.
\newblock {\em Proceedings of the Combustion Institute}, 32(1):1--25, 2009.

\bibitem{MSmodelingcomp}
E.~Weinan and B.~Engquist.
\newblock Multiscale modeling and computation.
\newblock {\em Notices of the American Mathematical Society}, 9(50):1062--1070, 2003.

\bibitem{POPE20131}
Stephen~B. Pope.
\newblock Small scales, many species and the manifold challenges of turbulent combustion.
\newblock {\em Proceedings of the Combustion Institute}, 34(1):1--31, 2013.

\bibitem{KennyScitech}
Kenneth~L. Budzinski and Paul~E. DesJardin.
\newblock On the development of coupled radiative flamelet generated manifolds to predict solid fuel flame spread in microgravity.
\newblock In {\em AIAA SciTech 2023}. AIAA-2023-0783, National Harbor, MD, January 2023.

\bibitem{Amolscitech}
Amol Salunkhe, Georgios Georgalis, Abani Patra, and Varun Chandola.
\newblock An ensemble-based deep framework for estimating thermo-chemical state variables from flamelet generated manifolds.
\newblock In {\em AIAA SciTech 2023}. AIAA-2023-1285, National Harbor, MD, January 2023.

\bibitem{MUELLER2020287}
Michael~E. Mueller.
\newblock Physically-derived reduced-order manifold-based modeling for multi-modal turbulent combustion.
\newblock {\em Combustion and Flame}, 214:287--305, 2020.

\bibitem{PERRY2022112286}
Bruce~A. Perry, Marc~T. {Henry de Frahan}, and Shashank Yellapantula.
\newblock Co-optimized machine-learned manifold models for large eddy simulation of turbulent combustion.
\newblock {\em Combustion and Flame}, 244:112286, 2022.

\bibitem{HAWORTH2010168}
D.C. Haworth.
\newblock Progress in probability density function methods for turbulent reacting flows.
\newblock {\em Progress in Energy and Combustion Science}, 36(2):168--259, 2010.

\bibitem{PEI20152006}
Yuanjiang Pei, Evatt~R. Hawkes, Sanghoon Kook, Graham~M. Goldin, and Tianfeng Lu.
\newblock Modelling n-dodecane spray and combustion with the transported probability density function method.
\newblock {\em Combustion and Flame}, 162(5):2006--2019, 2015.

\bibitem{DESJARDIN1996343}
Paul~E. DesJardin and Steven~H. Frankel.
\newblock Assessment of turbulent combustion submodels using the linear eddy model.
\newblock {\em Combustion and Flame}, 104(3):343--357, 1996.

\bibitem{SHIRIAN2023106046}
Yasaman Shirian, Jeremy~A.K. Horwitz, and Ali Mani.
\newblock On the convergence of statistics in simulations of stationary incompressible turbulent flows.
\newblock {\em Computers \& Fluids}, 266:106046, 2023.

\bibitem{JANSSEN2013123}
Hans Janssen.
\newblock Monte-carlo based uncertainty analysis: Sampling efficiency and sampling convergence.
\newblock {\em Reliability Engineering \& System Safety}, 109:123--132, 2013.

\bibitem{grammacytext}
Jiangeng Huang, Robert~B. Gramacy, Micka{\"e}l Binois, and Mirko Libraschi.
\newblock On-site surrogates for large-scale calibration.
\newblock {\em Applied Stochastic Models in Business and Industry}, 36(2):283--304, 2020.

\bibitem{KennedyOhagan}
MC~Kennedy and A~O'Hagan.
\newblock {Predicting the Output from a Complex Computer Code When Fast Approximations Are Available}.
\newblock {\em Biometrika}, 87(1):1--13, 03 2000.

\bibitem{sackswelch}
William~J. Welch, Robert.~J. Buck, Jerome Sacks, Henry~P. Wynn, Toby~J. Mitchell, and Max~D. Morris.
\newblock Screening, predicting, and computer experiments.
\newblock {\em Technometrics}, 34(1):15--25, 1992.

\bibitem{PCElit}
Lukas Novak and Drahomir Novak.
\newblock Polynomial chaos expansion for surrogate modelling: Theory and software.
\newblock {\em Beton- und Stahlbetonbau}, 113(S2):27--32, 2018.

\bibitem{LRAlit}
M.~Chevreuil, R.~Lebrun, A.~Nouy, and P.~Rai.
\newblock A least-squares method for sparse low rank approximation of multivariate functions.
\newblock {\em Society for Industrial and Applied Mathematics/American Statistical Association Journal on Uncertainty Quantification}, 3(1):897--921, 2015.

\bibitem{CHAN2018493}
Shing Chan and Ahmed~H. Elsheikh.
\newblock A machine learning approach for efficient uncertainty quantification using multiscale methods.
\newblock {\em Journal of Computational Physics}, 354:493--511, 2018.

\bibitem{singh2024framework}
Pratyush~Kumar Singh, Kathryn~A Farrell-Maupin, and Danial Faghihi.
\newblock A framework for strategic discovery of credible neural network surrogate models under uncertainty.
\newblock {\em Computer Methods in Applied Mechanics and Engineering}, 427:117061, 2024.

\bibitem{KrigingGP}
Jay~D. Martin and Timothy~W. Simpson.
\newblock Use of kriging models to approximate deterministic computer models.
\newblock {\em AIAA Journal}, 43(4):853--863, 2005.

\bibitem{RADAIDEH2020106731}
Majdi~I. Radaideh and Tomasz Kozlowski.
\newblock Surrogate modeling of advanced computer simulations using deep gaussian processes.
\newblock {\em Reliability Engineering \& System Safety}, 195:106731, 2020.

\bibitem{10.1007/11494669_93}
Michel Verleysen and Damien Fran{\c{c}}ois.
\newblock The curse of dimensionality in data mining and time series prediction.
\newblock In {\em Computational Intelligence and Bioinspired Systems}, pages 758--770, Berlin, Heidelberg, 2005. Springer.

\bibitem{hou2022dimensionality}
C.K.J. Hou and K.~Behdinan.
\newblock Dimensionality reduction in surrogate modeling: A review of combined methods.
\newblock {\em Data Science and Engineering}, 7:402--427, 2022.

\bibitem{10.5555/927743}
Carl~Edward Rasmussen.
\newblock {\em Evaluation of Gaussian Processes and Other Methods for Non-linear Regression}.
\newblock PhD thesis, 1997.

\bibitem{BUDZINSKI2020248}
Kenneth Budzinski, Siddhant~S. Aphale, Elektra~Katz Ismael, Gabriel Surina, and Paul~E. DesJardin.
\newblock Radiation heat transfer in ablating boundary layer combustion theory used for hybrid rocket motor analysis.
\newblock {\em Combustion and Flame}, 217:248--261, 2020.

\bibitem{petsc-user-ref}
Satish Balay, Shrirang Abhyankar, Mark~F. Adams, Steven Benson, Jed Brown, Peter Brune, Kris Buschelman, Emil Constantinescu, Lisandro Dalcin, Alp Dener, Victor Eijkhout, William~D. Gropp, V\'{a}clav Hapla, Tobin Isaac, Pierre Jolivet, Dmitry Karpeev, Dinesh Kaushik, Matthew~G. Knepley, Fande Kong, Scott Kruger, Dave~A. May, Lois~Curfman McInnes, Richard~Tran Mills, Lawrence Mitchell, Todd Munson, Jose~E. Roman, Karl Rupp, Patrick Sanan, Jason Sarich, Barry~F. Smith, Stefano Zampini, Hong Zhang, Hong Zhang, and Junchao Zhang.
\newblock {PETSc/TAO} users manual.
\newblock Technical report, ANL-21/39 - Revision 3.16, Argonne National Laboratory, 2021.

\bibitem{Shekhar2020b}
Prashant Shekhar and Abani Patra.
\newblock Hierarchical approximations for data reduction and learning at multiple scales.
\newblock {\em Foundations of Data Science}, 2(2):123--154, 2020.

\bibitem{Zilliac_Karabeyoglu_AIAA_RR}
G.~Zilliac and M.~Karabeyoglu.
\newblock Hybrid rocket fuel regression rate data and modeling.
\newblock {\em 42nd AIAA Joint Propulsion Conference and Exhibit}, AIAA 2006-4504, 2006.

\bibitem{Zilliac-uq}
Greg Zilliac, George~T. Story, Ashley~C. Karp, Elizabeth~T. Jens, and George Whittinghill.
\newblock Combustion efficiency in single port hybrid rocket engines.
\newblock {\em AIAA Propulsion and Energy 2020 Forum}, AIAA2020-3746, 2020.

\bibitem{NASA7}
Bonnie McBride, Michael Zehe, and Gordon Sanford.
\newblock Nasa glenn coefficients for calculating thermodynamic properties of individual species.
\newblock Technical Report NASA/TP 2002-211556, National Aeronautics and Space Administration (NASA), September 2002.

\bibitem{BCS21}
TA~Bolshova, AA~Chernov, and AG~Shmakov.
\newblock Reduced chemical kinetic mechanism for the oxidation of methyl methacrylate in flames at atmospheric pressure.
\newblock {\em Combustion, Explosion, and Shock Waves}, 57:159--170, 2021.

\bibitem{Turns00}
S.R. Turns.
\newblock {\em An Introduction to Combustion}.
\newblock McGraw Hill, New York, NY, 2000.

\bibitem{cantera}
David~G. Goodwin, Harry~K. Moffat, Ingmar Schoegl, Raymond~L. Speth, and Bryan~W. Weber.
\newblock Cantera: An object-oriented software toolkit for chemical kinetics, thermodynamics, and transport processes.
\newblock \url{https://www.cantera.org}, 2023.
\newblock Version 3.0.0.

\bibitem{chemkin}
R.~J. Kee, F.~M. Rupley, and J.~A. Miller.
\newblock {CHEMKIN-II}: A fortran chemical kinetics package for the analysis of gas-phase chemical kinetics.
\newblock Technical Report SAND-89-8009, Sandia National Laboratories, September 1989.

\bibitem{ABSD21}
Siddhant~S. Aphale, Kenneth Budzinski, Gabriel Surina, and Paul~E. DesJardin.
\newblock Influence of o2/n2 oxidizer blends on soot formation and radiative heat flux in pmma-air 2d slab burner for understanding hybrid rocket combustion.
\newblock {\em Combustion and Flame}, 234, 2021.

\bibitem{Modest22}
Michael~F. Modest and Sandip Mazumder.
\newblock {\em Radiation in Chemically Reacting Systems}, page 819–858.
\newblock Elsevier, 2022.

\bibitem{Zimmer16}
Leonardo Zimmer.
\newblock {\em Numerical Study of Soot Formation in Laminar Ethylene Diffusion Flames}.
\newblock PhD thesis, Universidade Federal do Rio Grande do Sul, 2016.

\bibitem{POINSOT1992104}
T.J Poinsot and S.K Lelef.
\newblock Boundary conditions for direct simulations of compressible viscous flows.
\newblock {\em Journal of Computational Physics}, 101(1):104--129, 1992.

\bibitem{KnepleyLG15}
Matthew~G. Knepley, Michael Lange, and Gerard~J. Gorman.
\newblock Unstructured overlapping mesh distribution in parallel.
\newblock {\em ArXiv}, arXiv:1506.06194, 2015.

\bibitem{kim:tchem:2021}
Kyungjoo Kim, Oscar Diaz-Ibarra, Habib~N. Najm, and Cosmin Safta.
\newblock {TChem v3.0} - a software toolkit for the analysis of complex kinetic models.
\newblock Technical Report SAND2021-14064, Sandia National Laboratories, Albuquerque, NM, 2021.

\bibitem{SHEKHAR2022115760}
Prashant Shekhar and Abani Patra.
\newblock Reprint of: A forward--backward greedy approach for sparse multiscale learning.
\newblock {\em Computer Methods in Applied Mechanics and Engineering}, 402:115760, 2022.

\bibitem{BERMANIS2013multiscale}
Amit Bermanis, Amir Averbuch, and Ronald~R. Coifman.
\newblock Multiscale data sampling and function extension.
\newblock {\em Applied and Computational Harmonic Analysis}, 34(1):15--29, 2013.

\bibitem{FLOATER199665}
Michael~S. Floater and Armin Iske.
\newblock Multistep scattered data interpolation using compactly supported radial basis functions.
\newblock {\em Journal of Computational and Applied Mathematics}, 73(1):65--78, 1996.

\bibitem{SHEKHAR2016887}
Prashant Shekhar, Abani Patra, and E.R. Stefanescu.
\newblock Multilevel methods for sparse representation of topographical data.
\newblock {\em Procedia Computer Science}, 80:887--896, 2016.

\bibitem{SHEKHAR20171652}
Prashant Shekhar, Abani Patra, and Beata~M. Csatho.
\newblock Multiscale and multiresolution methods for sparse representation of large datasets.
\newblock {\em Procedia Computer Science}, 108:1652--1661, 2017.

\bibitem{ZHANG2021337}
Zelin Zhang and Zhijun Wei.
\newblock Experiment and simulation of the effects of non-uniform magnetic field on the regression rate of pmma.
\newblock {\em Combustion and Flame}, 223:337--348, 2021.

\bibitem{gao2020uncertainty}
Connie~W Gao, Mengjie Liu, and William~H Green.
\newblock Uncertainty analysis of correlated parameters in automated reaction mechanism generation.
\newblock {\em International Journal of Chemical Kinetics}, 52(4):266--282, 2020.

\bibitem{zador2006local}
Judit Z{\'a}dor, I~Gy Zsely, and Tam{\'a}s Tur{\'a}nyi.
\newblock Local and global uncertainty analysis of complex chemical kinetic systems.
\newblock {\em Reliability Engineering \& System Safety}, 91(10-11):1232--1240, 2006.

\bibitem{ziehn2008global}
Tilo Ziehn and Alison~S Tomlin.
\newblock A global sensitivity study of sulfur chemistry in a premixed methane flame model using hdmr.
\newblock {\em International Journal of Chemical Kinetics}, 40(11):742--753, 2008.

\bibitem{sobol2007}
IM~Sobol’.
\newblock Global sensitivity analysis indices for the investigation of nonlinear mathematical models.
\newblock {\em Matematicheskoe Modelirovanie}, 19(11):23--24, 2007.

\bibitem{MIAO2016919}
Jianyu Miao and Lingfeng Niu.
\newblock A survey on feature selection.
\newblock {\em Procedia Computer Science}, 91:919--926, 2016.

\bibitem{GURURAJAN2019478}
Vyaas Gururajan and Fokion~N. Egolfopoulos.
\newblock Direct sensitivity analysis for ignition delay times.
\newblock {\em Combustion and Flame}, 209:478--480, 2019.

\bibitem{prager2013uncertainty}
Jens Prager, Habib~N Najm, Khachik Sargsyan, Cosmin Safta, and William~J Pitz.
\newblock Uncertainty quantification of reaction mechanisms accounting for correlations introduced by rate rules and fitted arrhenius parameters.
\newblock {\em Combustion and Flame}, 160(9):1583--1593, 2013.

\bibitem{sutton2016effects}
Jonathan~E Sutton, Wei Guo, Markos~A Katsoulakis, and Dionisios~G Vlachos.
\newblock Effects of correlated parameters and uncertainty in electronic-structure-based chemical kinetic modelling.
\newblock {\em Nature Chemistry}, 8(4):331, 2016.

\bibitem{saltelli2002making}
Andrea Saltelli.
\newblock Making best use of model evaluations to compute sensitivity indices.
\newblock {\em Computer Physics Communications}, 145(2):280--297, 2002.

\bibitem{dakota2020}
Brian~M. Adams, William~J. Bohnhoff, Keith~R. Dalbey, Mohamed~S. Ebeida, John~P. Eddy, Michael~S. Eldred, Russell~W. Hooper, Patricia~D. Hough, Kenneth~T. Hu, John~D. Jakeman, Mohammad Khalil, Kathryn~A. Maupin, Jason~A. Monschke, Elliott~M. Ridgway, Ahmad Rushdi, Daniel~Thomas Seidl, John~Adam Stephens, and Justin~G. Winokur.
\newblock Dakota, a multilevel parallel object-oriented framework for design optimization, parameter estimation, uncertainty quantification, and sensitivity analysis: Version 6.13 user's manual.
\newblock Technical Report SAND2020-12495, Sandia National Lab. (SNL-NM), Albuquerque, NM, 2020.

\bibitem{https://doi.org/10.1002/wics.1539}
Jiaxin Zhang.
\newblock Modern monte carlo methods for efficient uncertainty quantification and propagation: A survey.
\newblock {\em Wiley Interdisciplinary Reviews Computational Statistics}, 13(5):e1539, 2021.

\bibitem{oden2017predictive}
John~Tinsley Oden, Ivo Babu{\v{s}}ka, and Danial Faghihi.
\newblock Predictive computational science: Computer predictions in the presence of uncertainty.
\newblock {\em Encyclopedia of Computational Mechanics Second Edition}, pages 1--26, 2017.

\bibitem{tan2022predictive}
Jingye Tan, Pedram Maleki, Lu~An, Massimigliano Di~Luigi, Umberto Villa, Chi Zhou, Shenqiang Ren, and Danial Faghihi.
\newblock A predictive multiphase model of silica aerogels for building envelope insulations.
\newblock {\em Computational Mechanics}, 69(6):1457--1479, 2022.

\bibitem{liang2023bayesian}
Baoshan Liang, Jingye Tan, Luke Lozenski, David~A Hormuth, Thomas~E Yankeelov, Umberto Villa, and Danial Faghihi.
\newblock Bayesian inference of tissue heterogeneity for individualized prediction of glioma growth.
\newblock {\em Institute of Electrical and Electronics Engineers (IEEE) Transactions on Medical Imaging}, 42(10):2865--2875, 2023.

\bibitem{CV}
Richard~R. Picard and R.~Dennis Cook.
\newblock Cross-validation of regression models.
\newblock {\em Journal of the American Statistical Association}, 79(387):575--583, 1984.

\bibitem{SARI2014217}
Ahmet Sari, Cemil Alkan, and Cahit Bilgin.
\newblock Micro/nano encapsulation of some paraffin eutectic mixtures with poly(methyl methacrylate) shell: Preparation, characterization and latent heat thermal energy storage properties.
\newblock {\em Applied Energy}, 136:217--227, 2014.

\bibitem{Haario1999}
Heikki Haario, Eero Saksman, and Johanna Tamminen.
\newblock Adaptive proposal distribution for random walk metropolis algorithm.
\newblock {\em Computational Statistics}, 14(3):375–395, 1999.

\bibitem{nagy2011uncertainty}
Tibor Nagy and Tamas Turanyi.
\newblock Uncertainty of arrhenius parameters.
\newblock {\em International Journal of Chemical Kinetics}, 43(7):359--378, 2011.

\bibitem{medford2014assessing}
Andrew~J Medford, Jess Wellendorff, Aleksandra Vojvodic, Felix Studt, Frank Abild-Pedersen, Karsten~W Jacobsen, Thomas Bligaard, and Jens~K N{\o}rskov.
\newblock Assessing the reliability of calculated catalytic ammonia synthesis rates.
\newblock {\em Science}, 345(6193):197--200, 2014.

\end{thebibliography}

\newpage
\section*{Supplemental Material: Surrogate Comparison Figures}
\begin{figure}
    \centering
    \includegraphics[width = \textwidth]{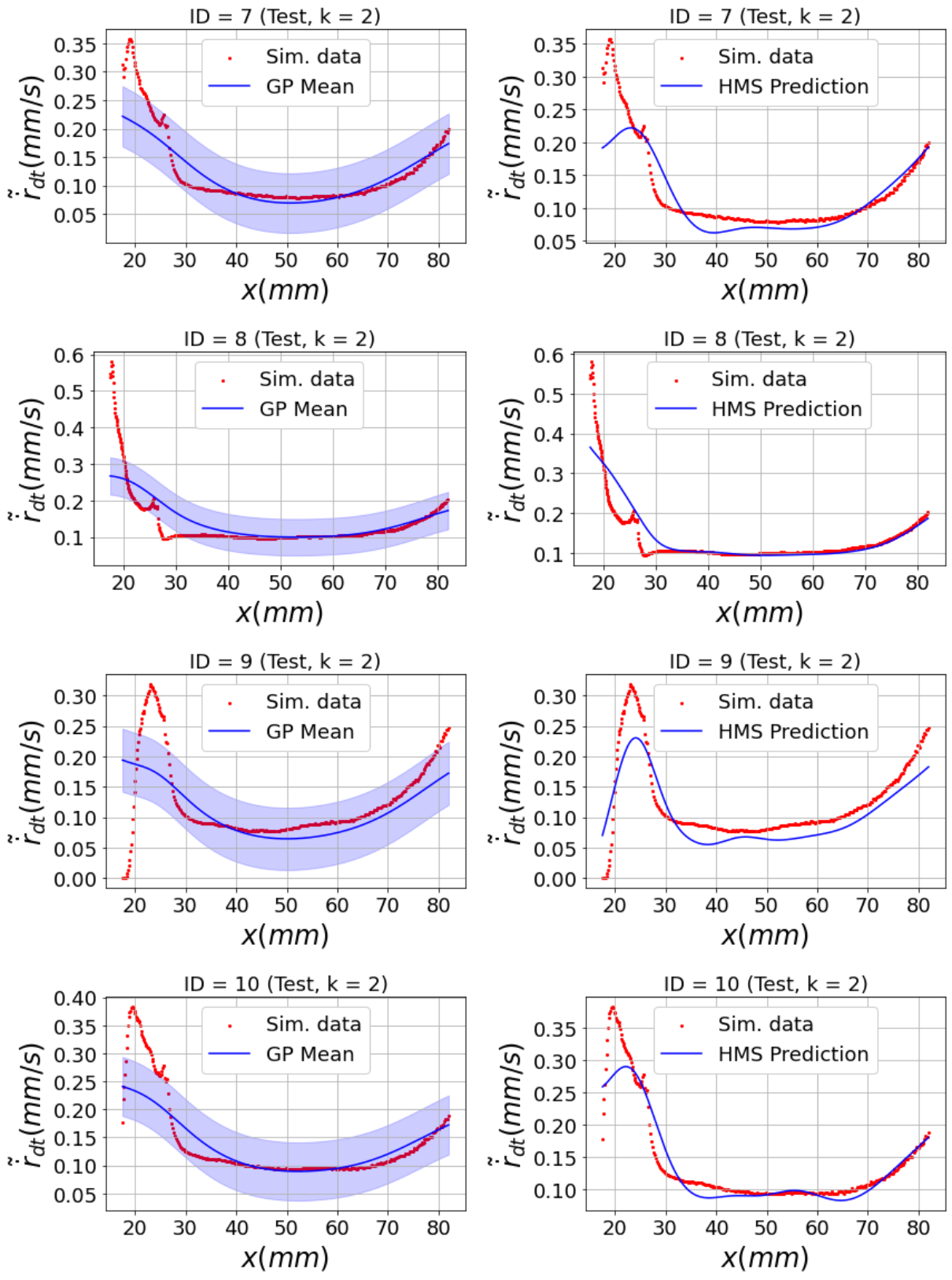}
    \caption{Additional comparisons between GP and HMS for other simulation IDs from the 5-fold cross validation. The conclusions are the same as we outlined in Section 4 of the main paper.}
    \label{fig:HMSvGP_3}
\end{figure}

\begin{figure}
    \centering
    \includegraphics[width = \textwidth]{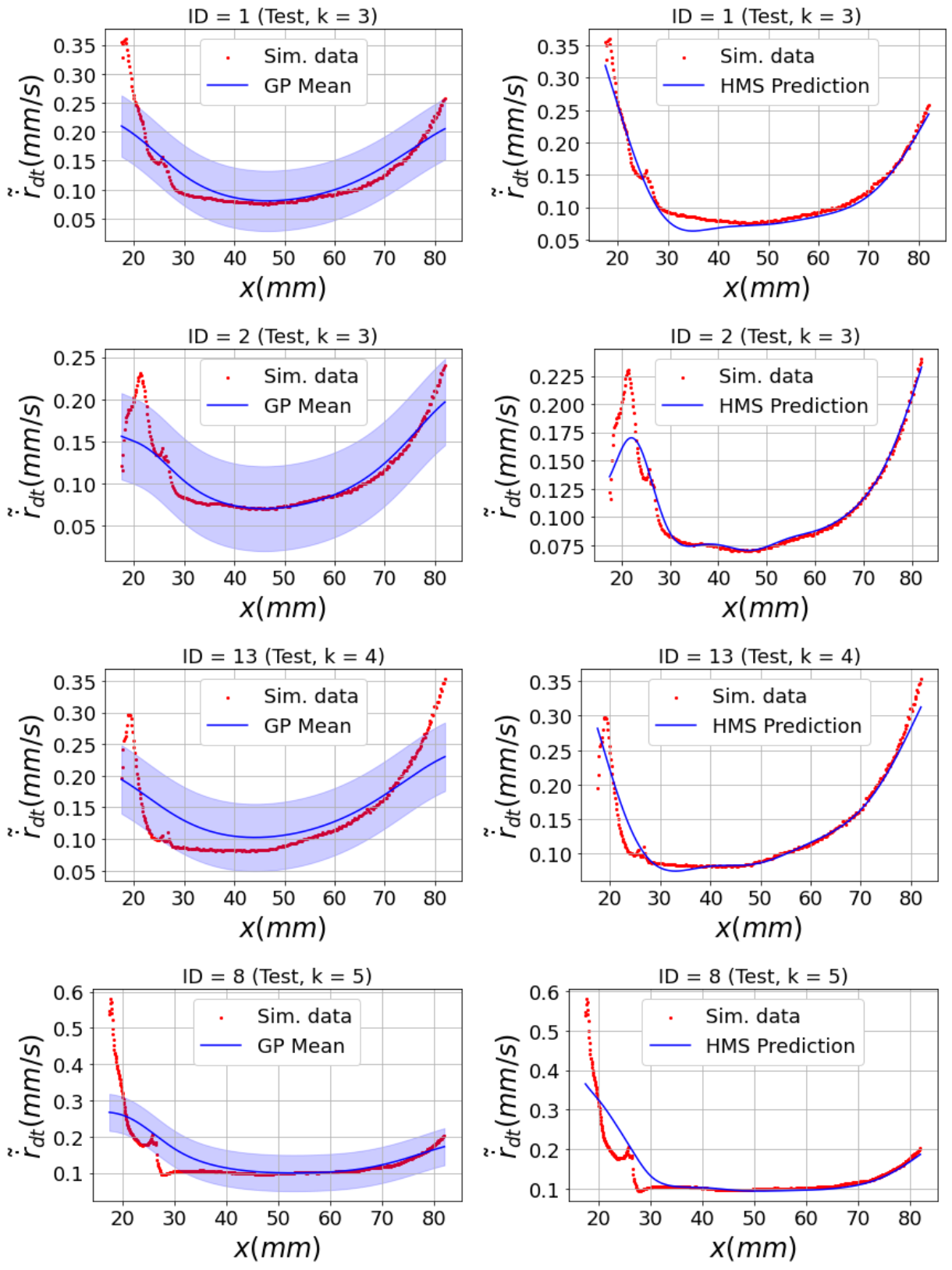}
    \caption{Additional comparisons between GP and HMS for other simulation IDs from the 5-fold cross validation. The conclusions are the same as we outlined in Section 4 of the main paper.}
    \label{fig:HMSvGP_4}
\end{figure}

Figs. \ref{fig:HMSvGP_3}, \ref{fig:HMSvGP_4} show additional comparisons between the GP and HMS for various simulation IDs across different folds.

\end{document}